\documentclass[12pt,preprint]{emulateapj}
\usepackage{natbib}

\slugcomment{To appear in ApJ}

\shorttitle{SONYC: NGC~1333}
\shortauthors{Scholz et al.}

\begin{document}
\bibliographystyle{apj}


\title{Substellar Objects in Nearby Young Clusters (SONYC): \\
The bottom of the Initial Mass Function in NGC~1333\footnote{Based on data collected at Subaru Telescope, which is operated by the National Astronomical Observatory of Japan.}}


\author{Alexander Scholz\altaffilmark{1}, Vincent Geers\altaffilmark{2}, 
Ray Jayawardhana\altaffilmark{2,**}, Laura Fissel\altaffilmark{2},
Eve Lee\altaffilmark{2}, David Lafreni{\`e}re\altaffilmark{2}, 
Motohide Tamura\altaffilmark{3}}

\email{as110@st-andrews.ac.uk}

\altaffiltext{1}{SUPA, School of Physics \& Astronomy, University of St. Andrews, North Haugh, St. Andrews, 
KY16 9SS, United Kingdom}
\altaffiltext{2}{Department of Astronomy \& Astrophysics, University of Toronto, 50 St. George Street, Toronto, 
ON M5S 3H4, Canada}
\altaffiltext{3}{National Astronomical Observatory, Osawa 2-21-2, Mitaka, Tokyo 181, Japan}
\altaffiltext{**}{Principal Investigator of SONYC}

\begin{abstract}
SONYC -- {\it Substellar Objects in Nearby Young Clusters} -- is a survey program to investigate the frequency and properties of substellar objects with masses down to a few times that of Jupiter in nearby star-forming regions. Here we present the first results from SONYC observations of NGC~1333, a $\sim 1$\,Myr old cluster in the Perseus star-forming complex. We have carried out extremely deep optical and near-infrared imaging in four bands (i', z', J, K) using Suprime-Cam and MOIRCS instruments at the Subaru telescope. The survey covers 0.25\,sqdeg and reaches completeness limits of 24.7\,mag in the i'-band and 20.8\,mag in the J-band. We select 196 candidates with colors as expected for young, very low-mass objects. Follow-up multi-object spectroscopy with MOIRCS is presented for 53 objects. We confirm 19 objects as likely brown dwarfs in NGC~1333, seven of them previously known. Nine additional objects are classified as possible stellar cluster members, likely with early to mid M spectral types. The confirmed objects are strongly clustered around the peak in the gas distribution and the core of the cluster of known stellar members. For 11 of them, we confirm the presence of disks based on Spitzer/IRAC photometry. The effective temperatures for the brown dwarf sample range from 2500\,K to 3000\,K, which translates to masses of $\sim 0.015$ to 0.1$\,M_{\odot}$, based on model evolutionary tracks. For comparison, the completeness limit of our survey translates to mass limits of 0.004$\,M_{\odot}$ for $A_V\la 5$\,mag or 0.008$\,M_{\odot}$ for $A_V\la 10$\,mag. Compared with other star-forming regions, NGC~1333 shows an overabundance of brown dwarfs relative to low-mass stars, by a factor of 2-5. On the other hand, NGC~1333 has a deficit of planetary-mass objects: Based on the surveys in $\sigma$\,Orionis, the Orion Nebula Cluster and Chamaeleon~I, the expected number of planetary-mass objects in NGC~1333 is 8-10, but we find none. It is plausible that our survey has detected the minimum mass limit for star formation in this particular cluster, at around 0.012-0.02$\,M_{\odot}$. If confirmed, our findings point to significant regional/environmental differences in the number of brown dwarfs and the minimum mass of the Initial Mass Function.
\end{abstract}


\section{Introduction}
\label{s1}

The origin of the stellar Initial Mass Function (IMF) is one of the major issues in astrophysics. The low-mass end of the IMF, in particular, has been the subject of numerous observational and theoretical studies over the past decade \citep[see][]{2007prpl.conf..149B}. While higher mass stars exhibit the well-known Salpeter shaped mass function  \citep[$dN \propto m^{-\alpha} dm$, with $\alpha = 2.35$,][]{1955ApJ...121..161S}, the power-law slope becomes significantly shallower below $\sim 0.5\,M_{\odot}$ \citep{2001MNRAS.322..231K,2003PASP..115..763C}. In the substellar regime, the coefficient $\alpha$ is mostly found to be around 0.6 \citep[e.g.][]{2003A&A...400..891M,2004A&A...416..555L,2007A&A...471..499M,2007A&A...470..903C}. While the higher-mass domain is thought to be mostly formed through fragmentation \citep[e.g.][]{2002ApJ...576..870P} and/or accretion onto the protostellar core \citep[e.g.]{2006MNRAS.370..488B}, in the low-mass and substellar regime additional physics is likely to play an important role, e.g. dynamical interactions \citep{2003MNRAS.339..577B,2009MNRAS.392..590B}, turbulence \citep{2004ApJ...617..559P}, photoerosion in the radiation field of bright stars \citep{2004A&A...427..299W}, or tidal shear in the gravitational potential of a stellar cluster \citep{2008MNRAS.389.1556B}.

To continue collapsing, an object or region has to cool as quickly as it is being heated by the conversion of gravitational into thermal energy. As radiative cooling is related to the opacity, this condition defines the opacity limit of fragmentation, which translates into a minimum mass for star-like sources, predicted to be between 0.001 and 0.01$\,M_{\odot}$ \citep[e.g.][]{1976MNRAS.176..367L,1976MNRAS.176..483R,2001ApJ...551L.167B,2005MNRAS.363..363B,2006A&A...458..817W}. Thus, isolated substellar objects in star-forming regions can have planetary-like masses. Furthermore, current numerical simulations of star formation build a significant fraction of brown dwarfs (BDs) by fragmentation in protostellar disks and subsequent ejection from the disk \citep{2003MNRAS.339..577B,2009MNRAS.392..413S}. Thus, a fraction of the substellar population may have a similar origin to massive planets. Hence, using the term `isolated planetary-mass object' \citep[IPMOs or planemos, see][]{2006AREPS..34..193B} for objects with masses below the Deuterium burning limit ($\sim 0.015\,M_{\odot}$) makes some physical sense, as this is the mass regime where star and planet formation overlap. 

The frequency and physical properties of the objects at the extreme low-mass end of the spectrum are poorly known. Following the discovery of brown dwarfs \citep{1995Natur.377..129R,1995Natur.378..463N}, surveys of star-forming regions, young clusters, and the field have revealed a rich population of substellar objects, triggering a wealth of follow-up studies to characterize their physics \citep[see][]{2000ARA&A..38..485B}. Populations of planetary-mass objects have been identified with masses down to $0.005-0.01\,M_{\odot}$ \citep[e.g.][]{2000Sci...290..103Z,2000MNRAS.314..858L}. So far, there is no evidence for a cut-off in the mass function, or in other words, the opacity limit has not been observed yet. 

As of today, however, only two regions have been systematically surveyed down to masses significantly below the Deuterium burning limit: $\sigma$\,Orionis cluster and the Orion Nebula Cluster (ONC). At a distance of $\sim 400$\,pc, any follow-up for planemos in Orion beyond photometry and very low-resolution spectroscopy is challenging. A few more objects in the same mass regime have been identified based on mid-infrared detections from ISO or Spitzer \citep{2002ApJ...571L.155T,2005ApJ...635L..93L,2006ApJ...644..364A,2008ApJ...684..654L}. This approach, however, will by definition only find sources with circumstellar material and thus yield incomplete results. The unbiased census for the members of the Chamaeleon I (ChaI) star-forming region is currently complete down to $\sim 0.01\,M_{\odot}$ \citep{2007ApJS..173..104L}, and has yielded a handful of planetary-mass members. 

SONYC -- Substellar Objects in Nearby Young Clusters -- is a new, unbiased attempt to establish the frequency of low-mass brown dwarfs and planetary-mass objects as a function of cluster environment and to provide the groundwork for detailed characterization of their physical properties (disks, binarity, atmospheres, accretion, activity). The project is based on extremely deep optical and near-infrared imaging and follow-up spectroscopy using 8-m class telescopes, aiming to detect the photospheres of the objects, which is a prerequisite for an unbiased survey. The observations are designed to reach limiting masses of 0.003$\,M_{\odot}$, and thus are meant to probe the opacity limit of star formation (see above). In each region, we aim to cover at least $\sim 1000$\,arcmin$^2$.

In the first paper for this project, we report on the survey in the cluster NGC~1333, part of the Perseus star-forming complex, carried out with Suprime-Cam and MOIRCS at the Subaru telescope. With an age of $\sim 1$\,Myr, a distance of 300\,pc \citep{1999AJ....117..354D,2002A&A...387..117B}, and moderate extinction, it is feasible to reach completeness limits of $I=25$\,mag and $J=21$\,mag, corresponding to masses for cluster members of about 0.003$\,M_{\odot}$, according to the COND03 evolutionary tracks \citep{2003A&A...402..701B}. Furthermore, the cluster is compact and can thus be covered efficiently. 

Near-infrared surveys in NGC~1333 have been carried out, among others, by \citet{1976AJ.....81..314S}, \citet{1994A&AS..106..165A}, and \citet{1996AJ....111.1964L}, establishing the presence of an embedded stellar cluster and a site of ongoing star formation with more than a hundred YSOs down to $K = 14.5$\,mag. More recently, \citet{2004AJ....127.1131W} have identified at least nine substellar objects based on near-infrared imaging over 133\,arcmin$^2$ with a limiting magnitude of $K=16$\,mag and follow-up spectroscopy. They find a disk frequency of 74\% and an upper limit to the mass function exponent of $\alpha \le 1.6$ in the substellar regime, consistent with previous studies. Building upon this work, \citet{2007AJ....133.1321G} have carried out deeper HST/NICMOS observations of a limited area (3\,arcmin$^2$), probing for the first time the planetary-mass regime. The near-infrared survey by \citet{2008AJ....136.1372O} is similarly deep ($K=18$\,mag) and covers 25\,arcmin$^2$, but does not include follow-up spectroscopy. In addition, NGC~1333 has been covered by numerous X-ray, mid-infrared, sub-millimeter, and millimeter surveys, for example in the framework of the Spitzer `Cores to Disks' Legacy Program, resulting in detailed studies of the cloud and cluster structure as well as accretion and outflow activity \citep[e.g.][]{2002ApJ...575..354G,2003A&A...401..543P,2006ApJ...638..293E,2006ApJ...645.1246J,2007ApJ...655..958W,2008ApJ...674..336G}.

This paper is structured as follows: We present our observations and data reduction procedures for photometry and spectroscopy in Sect.\ \ref{s2}. In Sect.\ \ref{s3} we discuss the photometric selection and spectroscopic verification of very low-mass objects in NGC~1333. The properties of our sources, including evidence for youth, are analysed in Sect.\ \ref{s4}. The low-mass end of the IMF based on our new findings is discussed in Sect.\ \ref{s5}. We give our conclusions in Sect.\ \ref{s6}.

\section{Observations and data reduction}
\label{s2}

\subsection{Optical imaging and photometry}
\label{s21}

We observed NGC~1333 in the SDSS i' and z' filters using the Subaru Prime Focus Camera (Suprime-Cam) wide field imager \citep{2002PASJ...54..833M} on 2006 November 17. The conditions were photometric with typical seeing ranging from 0.45-0.63\arcsec. Suprime-Cam is a mosaic camera which utilizes 10 CCDs arranged in a 5 by 2 pattern giving a total field of view of $34'\times27'$. As NGC~1333 is quite compact, the entire cluster was observed in a single pointing.  A ten point dither pattern was carried out to eliminate gaps between the CCDs and to correct for bad pixels. We observed this pattern in each filter six times, where the individual images have exposure times of 60\,sec, for a total integration time of 3600\,sec in each band.

A standard reduction was performed for each individual CCD chip. Routines from the Suprime-Cam data reduction software package {\em SDFRED} 
\citep{2002AJ....123...66Y,2004ApJ...611..660O} were used for overscan subtraction and flat fielding. In both filters, the images show significant extended structure from the molecular cloud emission. Sky subtraction was performed by fitting a Gaussian to the peak of a histogram of pixel brightnesses and subtracting the centroid.  Additional routines from {\em SDFRED} were used for distortion correction, bad pixel masking and image combination. Only one of the ten CCD chips  ({\tt w67c1}) showed noticeable fringing. To reduce each chip consistently, and because the 10 point dither pattern mostly removes the fringes, we did not carry out a fringe correction. The world coordinate system was calibrated against sources from the 2MASS point source catalog \citep{2003tmc..book.....C} with $J\ge 13$ using the {\em msctpeak} program from the IRAF package {\em MSCRED}\footnote{IRAF is distributed by the National Optical Astronomy Observatories, which are operated by the Association of Universities for Research in Astronomy, Inc., under cooperative agreement with the National Science Foundation.}. The typical fitting residuals were of order 0.25\arcsec.

Sources were identified from each CCD chip using the Source Extractor ({\em SExtractor}) software package \citep{1996A&AS..117..393B}. For object extraction we required at least 5 pixels to be above the 3$\sigma$ detection limit. {\em SExtractor} automatic aperture fitting photometry routines were used to calculate the flux of each source. We rejected objects within a few pixels of the edge of the image, elongated objects ($a/b>1.2$) and objects that were not present in both the i' and z' source lists. The rejection criteria were chosen conservatively, to minimize the number of spurious detections in the photometry database. We visually checked the detection algorithm and confirmed that the number of point-sources that are missed as well as the number of spurious sources that are included is minimal. The final optical catalogue has 7757 objects.

We measured the chip-to-chip zeropoint offsets from the median fluxes of domeflat images. The absolute zeropoint for the mosaic was derived from observations of the SA95 standard field, which contains SDSS secondary standards \citep{2002AJ....123.2121S}. To avoid saturation, the images of standard fields were defocused. The fluxes of five standard stars were extracted using aperture photometry with 50-pixel radius. We derived a mean zeropoint of $27.80$~$\pm$~$0.01$ in i'- and $27.12$~$\pm$~$0.02$ in z'-band. These values agree with the zeropoint measured by \citet{2002PASJ...54..833M} within $0.12$ magnitude in i' and $0.07$ in z'. The airmass difference between standard fields and science fields was $\pm 0.1$, i.e.\ the differential extinction is negligible. 

\begin{figure}
\center
\includegraphics[width=8.5cm]{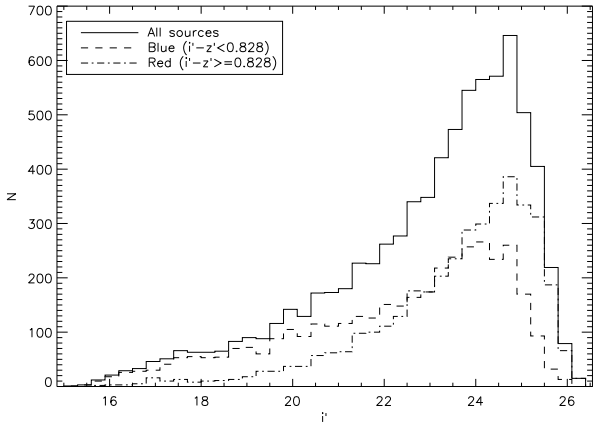} 
\caption{i'-band histogram of the objects in our photometric catalogue with i'- and z'-band data. The peak at 24.7\,mag indicates the completeness limit of the survey. The faintest objects in the survey are found at $i>26$\,mag. The dashed and dash-dotted lines show the histogram for equally sized samples of blue and red objects, illustrating that the completeness limit does not significantly change with color, although we are slightly more sensitive to red objects.
\label{f3}}
\end{figure}

From the histogram of the magnitudes (see Fig.\ \ref{f3} for the i'-band data) we derived completeness limits of our survey: In the i'-band, we are complete down to $24.7 \pm 0.2$\,mag, in z'-band down to $23.8\pm 0.2$\,mag. These limits are robust against changes in color, as demonstrated in Fig.\ \ref{f3}. The limiting magnitudes of our survey are about 26.5\,mag in i' and 25.5\,mag in z', which are close to the sensitivity limits determined during the Suprime-Cam commissioning phase \citep{2002PASJ...54..833M}.

\subsection{Near-infrared imaging and photometry}
\label{s22}

NGC~1333 was observed with the Multi-Object InfraRed Camera and Spectrograph (MOIRCS) mounted on the Subaru Telescope, in the MKO broad-band filters J and Ks, during the nights of 2008 December 4--6. MOIRCS uses two CCDs providing a total field of view of $4' \times 7'$ with a spatial resolution of 0.117 arcsec/pixel. NGC~1333 was covered in 24 pointings in both filters, using a 6 or 8-point dither pattern. In the J-band, the typical exposure time per pixel was 600\,sec, in the K-band 300\,sec.
In addition, the center of NGC~1333 was covered in a smaller series of 13\,sec short exposures, to obtain photometry for the brightest sources that saturate in the deeper exposures.

Data reduction was performed using a modified version of the SIMPLE-MOIRCS package\footnote{http://www.aoc.nrao.edu/$\sim$whwang/idl/SIMPLE/MOIRCS/Doc/}, which is written in IDL\footnote{Interactive Data Language} and uses SExtractor. Background subtraction, flatfield correction and distortion correction were performed on individual images, before the images were co-added. The world coordinate system was calibrated against the 2MASS point-source catalogue. Initial source identification is performed using SExtractor, requiring at least 3 pixels to be above the 3$\sigma$ detection limit. Source rejection was performed based on saturation and elongation. After running the identification and rejection routines, the sources were visually verified for each field. Photometry was performed using a fixed-aperture extraction routine in IDL. To account for variable seeing between fields, an aperture correction factor is determined for each individual field. To correct for difference in sensitivity between detector 1 and 2, a flux offset factor is determined from the average level of the sky flux in both images, and applied to detector 2 for each individual field. 

The final near-infrared catalogue has 2360 sources. Calibration of the absolute zeropoint magnitude is performed by comparing the full catalogue of extracted sources to the 2MASS point source catalogue. The mean zeropoint is derived to be 25.08 $\pm$ 0.03 in J-band and 24.3 $\pm$ 0.04 mag in K-band. Completeness limits in each band are derived from the number distribution of the magnitudes (see Sect. \ref{s21}). The imaging is complete down to 18.0\,mag in K-band and 20.8\,mag in J-band.

\subsection{Multi-object spectroscopy}
\label{s23}

We used MOIRCS again to carry out multi-object spectroscopy (MOS) for 53 sources in NGC~1333, 36 of them selected based on the photometry in the four bands i', z', J, K (see Sect. \ref{s3}). The targets are covered in six multi-slit masks; the number of objects per mask was 5-13. Pre-imaging for the masks in the K-band was obtained in September 2008. We used slits which are 0.9\arcsec~wide and 9-12\arcsec~long, except for a few longer slits, used to observe standard stars. The spectroscopy run was carried out in three nights on Dec 4-6, 2008. For each mask we integrated in total 50-60\,min with the grism HK500, split in shorter exposures of 5 or 10\,min, depending on conditions. Between the single exposures, the mask was moved so that the targets shifted by 2.5\arcsec~along the slit (nodding), to facilitate sky subtraction. The seeing during the science integrations was stable at 0.5-0.7\arcsec. Before and after the NGC~1333 masks, we observed A0 stars through one of the science masks for flux calibration and extinction correction; these exposures required de-focusing to prevent saturation. For each mask, we obtained series of domeflats with lamp on and off for calibration purposes.

The MOS data was reduced following standard recipes for near-infrared spectra. In a first step, we subtracted the pairs of nodded exposures, which removes the sky background and the detector bias. These images were divided by a normalized flatfield, which is the difference between the averaged lamp-on and lamp-off domeflats. Frames from the same nodding position were co-added. This gives us two final images for each chip and mask. In these frames, the signal-to-noise ratio for the spectra ranges from 20 for the faintest to 1000 for the brightest sources. These values have been measured in the central parts of the H- and K-band, in regions without strong skyline residuals.

For the extraction of the spectra we used the {\it apall} routine within IRAF. All objects from our photometric candidate list are well-detected. For wavelength calibration we additionally extracted the spectra from the unreduced frames, which contain plenty of telluric OH lines. As the wavelength solution depends slightly on the position of the slit on the chip, the fit was done for each object separately. For the HK grism, a second order polynomial fit covering about 20 skylines gave a typical rms of 2-3\,\AA, well below the resolution. All spectra are smoothed with a $\pm 25$\,\AA~boxcar average and binned to 40\,\AA~per pixel. The two nodded spectra for the same object were coadded. The standard star spectra were treated in exactly the same way as the science frames.

Near-infrared spectra show strong effects of the atmosphere -- broad absorption features, mostly between the J/H and H/K bands, as well as sharp OH emission lines. In our data, subtracting the nodding exposures reduced the strength of the OH lines by 95\% or better. The residuals are removed in the background subtraction which is part of the extraction. The corrections for telluric extinction and instrument response are usually done in one step, based on observations of a standard star with known spectral energy distribution. Our A0 standard star spectra were divided by a library spectrum of an A0 star \citep{1998PASP..110..863P}, obtained from the ESO website\footnote{http://www.eso.org/sci/facilities/paranal/instruments/isaac/tools/
lib/index.html}. To remove residuals of the Hydrogen absorption features seen in A0 spectra, we linearly interpolated from 1.54 to 1.78$\,\mu m$ and from 2.09 to 2.34$\,\mu m$, regions that are mostly free from additional telluric absorption bands.

All science spectra were divided by one of these modified standard spectra observed at similar airmass. Typically, the airmass difference is $<0.1$. Still, we find that the correction of atmospheric correction is not optimum. We attribute this mainly to the duration of our science exposures: While the standards were observed within a few minutes, the science exposures cover 1-2 hours. Variable sky conditions, as they have been present during the observations, hamper a good extinction correction. For our analysis we therefore focus on the spectral regions least affected by telluric absorption, i.e.\ 1.5-1.75$\,\mu m$ in the H-band and 2.1-2.3$\,\mu m$ in the K-band.

\section{Selection of new substellar members in NGC~1333}
\label{s3}

\subsection{Optical color-magnitude diagram}
\label{s31}

\begin{figure*}
\center
\includegraphics[width=16cm]{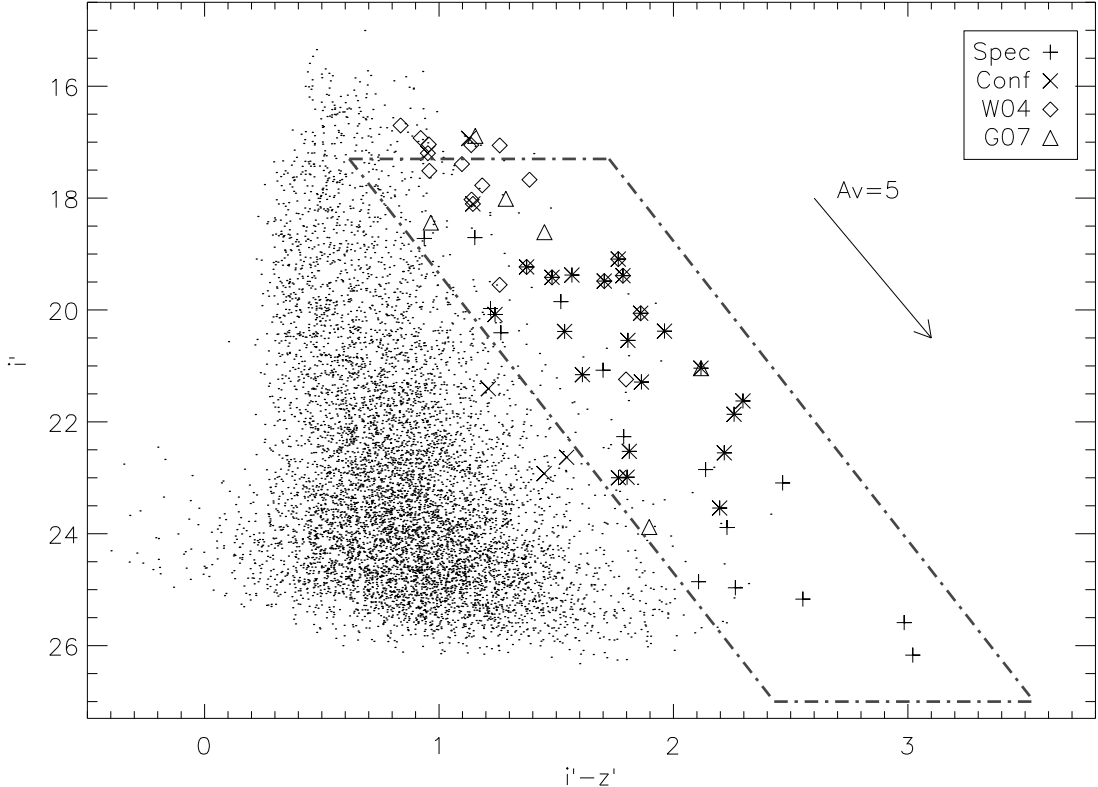} 
\caption{Color-magnitude diagram in (i', i'--z') constructed from our deep Suprime-Cam observations. The selection box for follow-up spectroscopy is shown with dash-dotted lines. Objects chosen for spectroscopy (plus) and confirmed or possible M-type members of NGC~1333 (cross, see Table \ref{t1} and \ref{t2}) are marked. Spectroscopically confirmed BDs from the surveys by \citet{2004AJ....127.1131W} and \citet{2007AJ....133.1321G} are marked with diamond and triangle symbols respectively.
\label{f2}}
\end{figure*}

The goal of the project is to identify new very low-mass members of NGC~1333. Due to their pre-main sequence status, plus extinction, these sources are expected to occupy a distinct area in optical/near-infrared color-magnitude diagrams, on the red side of the broad cumulation of background main sequence stars. In Fig.\ \ref{f2} we show the (i', i'--z') diagram, which we used to identify the initial list of candidates. The cumulation of non-members around $i'-z' = 1.0$\,mag is clearly seen. Previously identified very low-mass members of NGC~1333 with spectroscopic confirmation from the papers by \citet{2004AJ....127.1131W} and \citet{2007AJ....133.1321G} are marked. 

Based on the locus of known members, we construct a selection box for this diagram, which encompasses the color-magnitude range where we expect cluster members with lower masses to be found (dash-dotted lines). The position of the box is chosen based on the colors of known substellar cluster members in the diagram. Apart from this empirical constraint, we aimed to select candidates with a broad range in colors, to maximize our chances of covering the cluster population. In particular, the choice of the selection box is independent from models. At the bright end, the box is limited by the onset of saturation at  $i' \sim 17.3$\,mag. The box extends down to our limiting magnitude of $i' \sim 26.5$\,mag. The selection box contains 196 objects with reliable photometry in i'- and z'-bands. From this initial sample, we select a representative list of 36 sources for follow-up spectroscopy, marked with $+$-signs in Fig.\ \ref{f2}. This sample includes 19 objects from the sample from \citet{2004AJ....127.1131W}, six of them previously spectroscopically confirmed as members.

In the selection of candidates for spectroscopy, we ignore the bright end of the selection box $i' \lesssim 18.5$\,mag, which contains about one third of the candidates. Cluster members with such magnitudes are likely to have masses around or above the substellar limit, and are therefore of less interest to us. Apart from that, we aim to sample the full range of colors and magnitudes in the box, to account for the uncertainties in estimating the location of the cluster members in this diagram. Thus, the `spectroscopic' sample does not show any bias in terms of colors or magnitudes in comparison with the full photometric candidate sample. In addition, there is no bias in spatial distribution: the full photometric sample shows a clear clustering around the cloud cores in NGC~1333, thus our spectroscopy fields are also distributed in this area (see Sect \ref{s43}). Since the density of young sources is higher in the regions close to the cloud cores, the contamination with field objects is expected to be larger further away from the cores. Thus, focusing on the cores is reasonable for minimizing contamination as well.

\subsection{Near-infrared color-magnitude diagrams}
\label{s32}

\begin{figure*}
\center
\includegraphics[width=16cm]{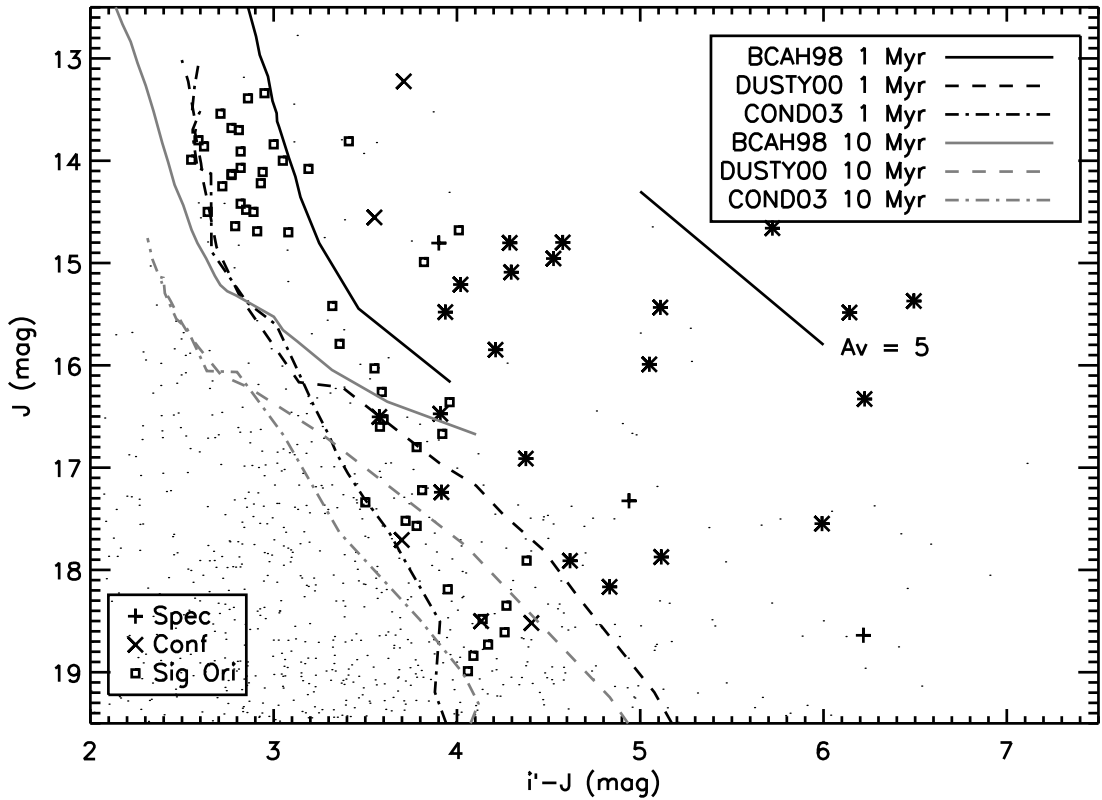} 
\caption{Color-magnitude diagram in (J, i'--J) constructed from our MOIRCS and Suprime-Cam observations. The candidates selected for spectroscopy from the optical diagram and the confirmed M-type objects (see Table \ref{t1} and \ref{t2}) are marked with plusses and crosses, respectively. For comparison, we overplot model isochrones for 1 and 10\,Myr and the photometry for the substellar objects in $\sigma$\,Ori \citep{2007A&A...470..903C}, shifted to the distance of NGC~1333. For the models and the $\sigma$\,Ori objects the Cousins-I photometry was shifted to the Sloan-i' band using the transformation given by \citet{2006A&A...460..339J}.
\label{f1}}
\end{figure*}

We used the near-infrared photometry to verify the initial candidate selection. In particular, we aim to ensure that our candidates show the  colors expected for reddened very low-mass members of NGC~1333. In Fig.\ \ref{f1} we show the color-magnitude diagram constructed by combining the Suprime-Cam i'-band photometry with the MOIRCS J-band data. The diagram shows all objects with data in both bands, but objects selected for spectroscopy and the 28 objects confirmed by the spectroscopy (Sect.\ \ref{s33}) are marked. A few of the confirmed objects are not in our near-infrared catalogue due to saturation; for them we use 2MASS data instead.

For comparison, we overplot the BCAH98, COND03, and DUSTY00 evolutionary tracks for 1 and 10\,Myr from \citet{1998A&A...337..403B,2000ApJ...542..464C,2003A&A...402..701B} and the photometry for the substellar population in the $\sigma$\,Ori cluster \citep{2007A&A...470..903C}. In both cases, we converted the Cousins I-band data to the Sloan i'-band using the transformation given by \citet{2006A&A...460..339J}. In principle this requires knowledge of the $R-I$ color, which is not always available for the $\sigma$\,Ori sources. We assumed an average $R-I = 2.0$ in these cases; the individual (i'--J) colors for $\sigma$\,Ori objects may therefore be incorrect by about $\pm 0.2$. Inconsistencies between the Suprime-Cam i'-band with the standard bandpass may cause additional uncertainties. The $\sigma$\,Ori photometry was shifted to the distance of NGC~1333, assuming a differential distance modulus of 0.7, consistent with recent estimates. $\sigma$\,Ori is a cluster with little extinction; we expect these objects to be mostly free from reddening. 

The plot demonstrates that the spectroscopy candidates cover the color space that is limited on the blue side by models and the $\sigma$\,Ori sources. This is as expected, since the cloud cores in NGC~1333 have significant and variable extinction so the population of young cluster members as well as background objects should be scattered towards the red side. Apart from reddening, there is no systematic offset between our sample and the comparison data. Thus, we confirm that our selection of candidates covers the color space expected for very low-mass members of NGC~1333.

\begin{figure}
\center
\includegraphics[angle=-90,width=8.5cm]{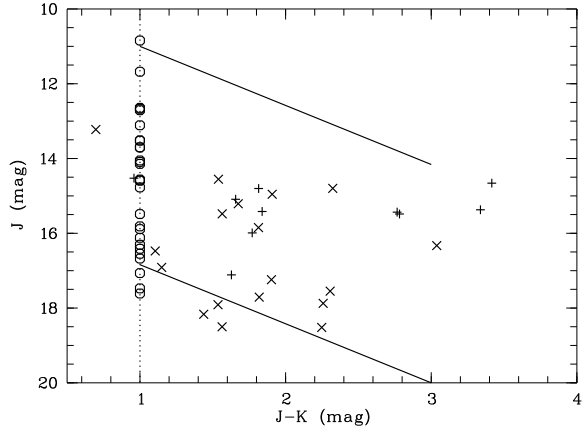} 
\caption{Near-infrared color-magnitude diagram for the 28 objects listed in Table \ref{t1} and \ref{t2}. The plusses show the MOIRCS photometry, crosses are 2MASS values for the cases not covered by our data, the circles the dereddened values, assuming intrinsic colors of $J - K = 1.0$\,mag (dotted line). The solid lines illustrate the reddening path.
\label{f7}}
\end{figure}

The intrinsic near-infrared color $J-K$ of late-type objects is mostly independent of spectral type and luminosity; thus it can be used to estimate the extinction for the suspected cluster members. This is done only for the objects confirmed as having spectral type M or later (Sect.\ \ref{s33}). We assume here an intrinsic color of $J-K = 1.0$\,mag. Within $\pm 0.2$\,mag, this is consistent with the empirical values for main sequence dwarfs and giants with spectral type M \citep{1988PASP..100.1134B}, with the predictions from the evolutionary tracks COND03 (for 0.003-0.1$\,M_{\odot}$), BCAH98 (for 0.03-0.5$\,M_{\odot}$), and DUSTY00 (for 0.006-0.1$\,M_{\odot}$), and with the colors of brown dwarfs in the 3\,Myr old $\sigma$\,Ori star-forming region down to 0.013$\,M_{\odot}$ \citep{2007A&A...470..903C}. For young objects with masses below the Deuterium burning limit there are some indications that the $J-K$ color increases to 1-1.5 \citep{2006MNRAS.373...95L,2007A&A...470..903C}; this may lead to an overestimate of the extinction by as much as 2.5\,mag for the lowest mass objects.

For the extinction estimate and throughout this paper, we use the extinction law from \citet{1990ARA&A..28...37M}, with $R_V = 4$. In Fig.\ \ref{f7} we plot the ($J$, $J-K$) color-magnitude diagram with the solid line indicating the reddening path and the dotted line showing the assumed intrinsic color. For the 28 objects with spectroscopic confirmation (listed in Tables \ref{t1} and \ref{t2}), we plot our photometry (crosses) or 2MASS data (plusses), as well as the dereddened J-band photometry (circles). For the optical extinction in our confirmed sample, we obtain $A_V = 0-13$, i.e.\ we are probing deeper into the cloud than \citet[][$A_V\le7$]{2004AJ....127.1131W}. The results for $A_V$ are given in Tables \ref{t1} and \ref{t2}. 

This range of optical extinctions is in good agreement with the extinction maps as derived from submillimeter continuum observations \citep{2006ApJ...638..293E}. Converting a 1.1\,mm map to an extinction map results in $A_V<17$\,mag for 5$'$ resolution for NGC~1333. Our extinction estimates for individual objects are all in this range, perhaps with the exception of object \#12 (see discussion in Sect. \ref{s33}). The comparison also indicates that we do not miss a significant number of additional objects at even higher extinctions. Note that the maximum extinction is lower when derived from 2MASS near-infrared data or from $^{13}$CO observations, mainly because these indicators do not trace the densest regions of the cloud well.

The main error sources in the extinction estimates are a) uncertainties in the estimate of the intrinsic colors and b) possible excess emission due to disks and accretion. As pointed out above, except for the lowest mass objects, the intrinsic colors are likely to be accurate within 0.2\,mag, which causes errors in $A_V$ of $<1$\,mag. At the lowest masses, the error may be larger. The disk fraction in NGC~1333 is high \citep[83\%,][]{2008ApJ...674..336G}, hence a substantial fraction of cluster members may show enhanced K-band flux due to excess emission from dust in the disk, which leads to a systematic overestimate of the extinction. On the other hand, accretion induced hot spots can enhance the J-band as well as the K-band flux \citep{1999A&A...352..517F}. Prior to a more detailed characterisation, these effects are difficult to take into account. We conservatively estimate that our typical uncertainty in $A_V$ is 2\,mag; larger errors are possible in individual cases. The extinction values will be used in the spectral fitting in Sect.\ \ref{s33}.

\subsection{Spectroscopy}
\label{s33}

With the multi-object spectroscopy described in Sect.\ \ref{s23} we obtained low-resolution spectra for 53 objects in total. The sample of objects with spectra includes 36 candidates from the primary photometric sample discussed in Sect.\ \ref{s31}. In addition, we have spectra for nine objects that are slightly bluer in optical colors than our primary candidates sources, to verify the validity of the optical color criterion (called `blue' objects in the following). Among our spectroscopy sample are nine confirmed young M dwarfs from the sample published by \citet{2004AJ....127.1131W}, six of them also in our own photometric candidate sample. Finally, we included five random objects in our MOS fields. We had 12 additional slits on objects that are not well detected and did not allow a reliable extraction. All these undetected sources have been included as either blue or random objects, and do not come from our photometric sample.

\begin{figure}
\center
\includegraphics[width=8.5cm]{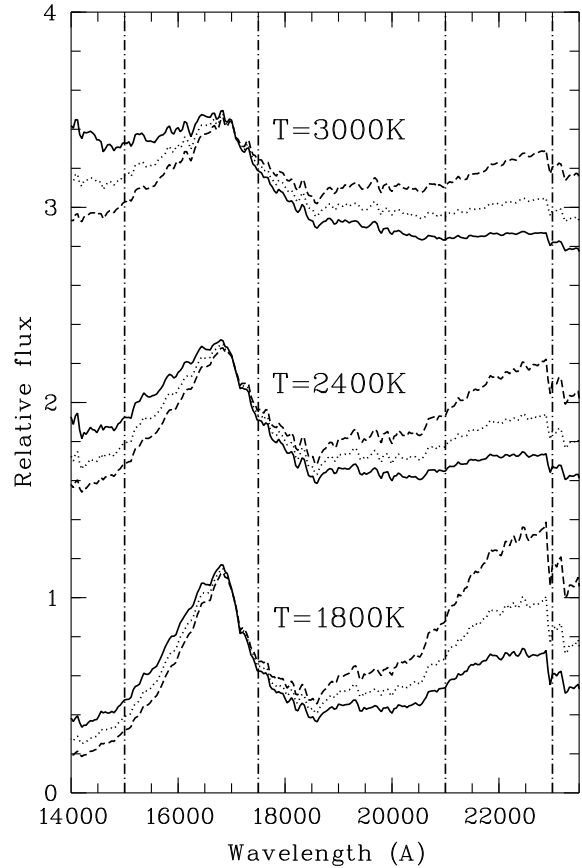} 
\caption{Reddened model spectra from the AMES DUSTY series \citep{2001ApJ...556..357A} for $\log{g}  = 3.5$, corresponding to pre-main sequence ages. Shown are three different temperatures (from top to bottom: 3000\,K, 2400\,K, 1800\,K), each with three different extinctions $A_V=0, 5, 10$\,mag as solid, dotted, dashed lines. The H-band peak and its dependence on temperature is clearly visible. The relative strength of H- and K-band is strongly affected by extinction. Dash-dotted lines show the spectral range in our observations used for classification.
\label{f4}}
\end{figure}

Our goal is to identify young members of NGC~1333 with substellar masses, i.e.\ with spectral types later than M5. In addition to these young brown dwarfs, our sample may contain embedded stellar members of NGC~1333 and reddened background stars with spectral types earlier than M5, and (less likely) late M and early L type field objects in the foreground. We show some examples of the expected shape of the spectra for reddened young brown dwarfs, derived from model spectra, in Fig.\ \ref{f4}.

\begin{figure*}
\center
\includegraphics[width=8.8cm]{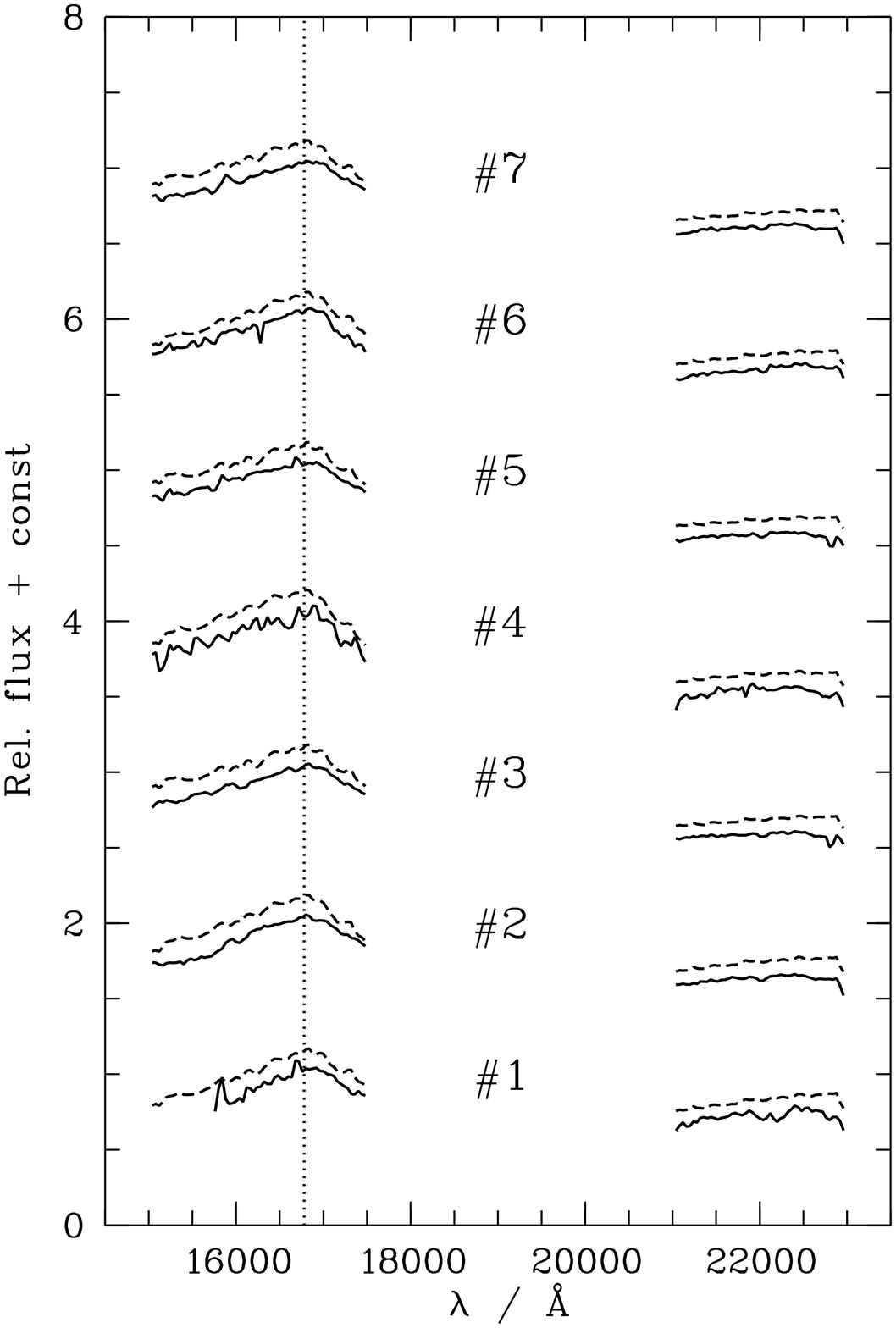} \hfill
\includegraphics[width=8.8cm]{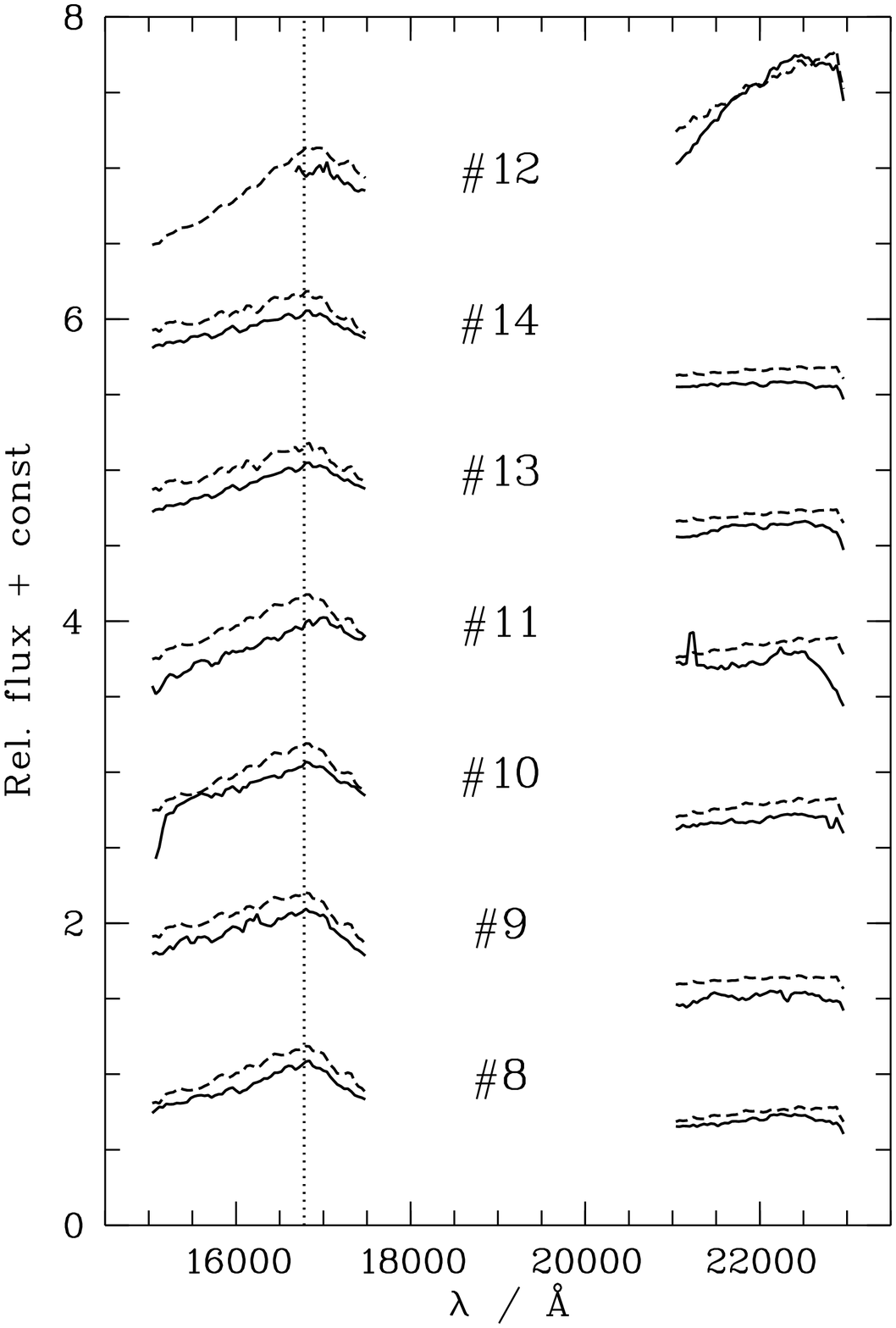} \\
\caption{Spectra of late-type objects in NGC~1333 (solid lines) in comparison with models (dashed lines), first part. The running number is the same as in Tables \ref{t1} and \ref{t2}. Shown is the best fit model with the effective temperature as given in the tables. Objects 1-19 are probable substellar members of NGC~1333.
\label{f5a}}
\end{figure*}

\begin{figure*}
\center
\includegraphics[width=8.8cm]{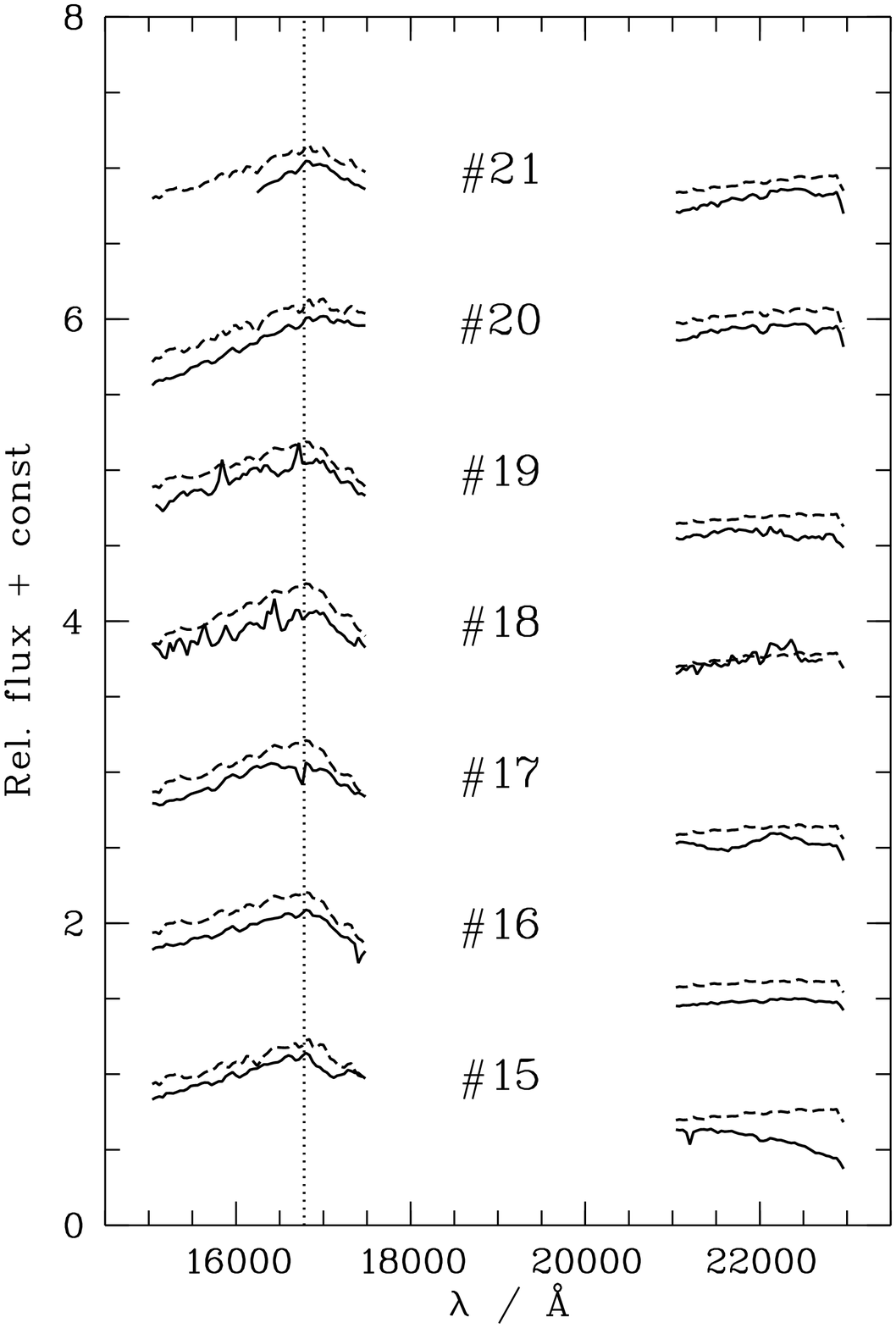} \hfill
\includegraphics[width=8.8cm]{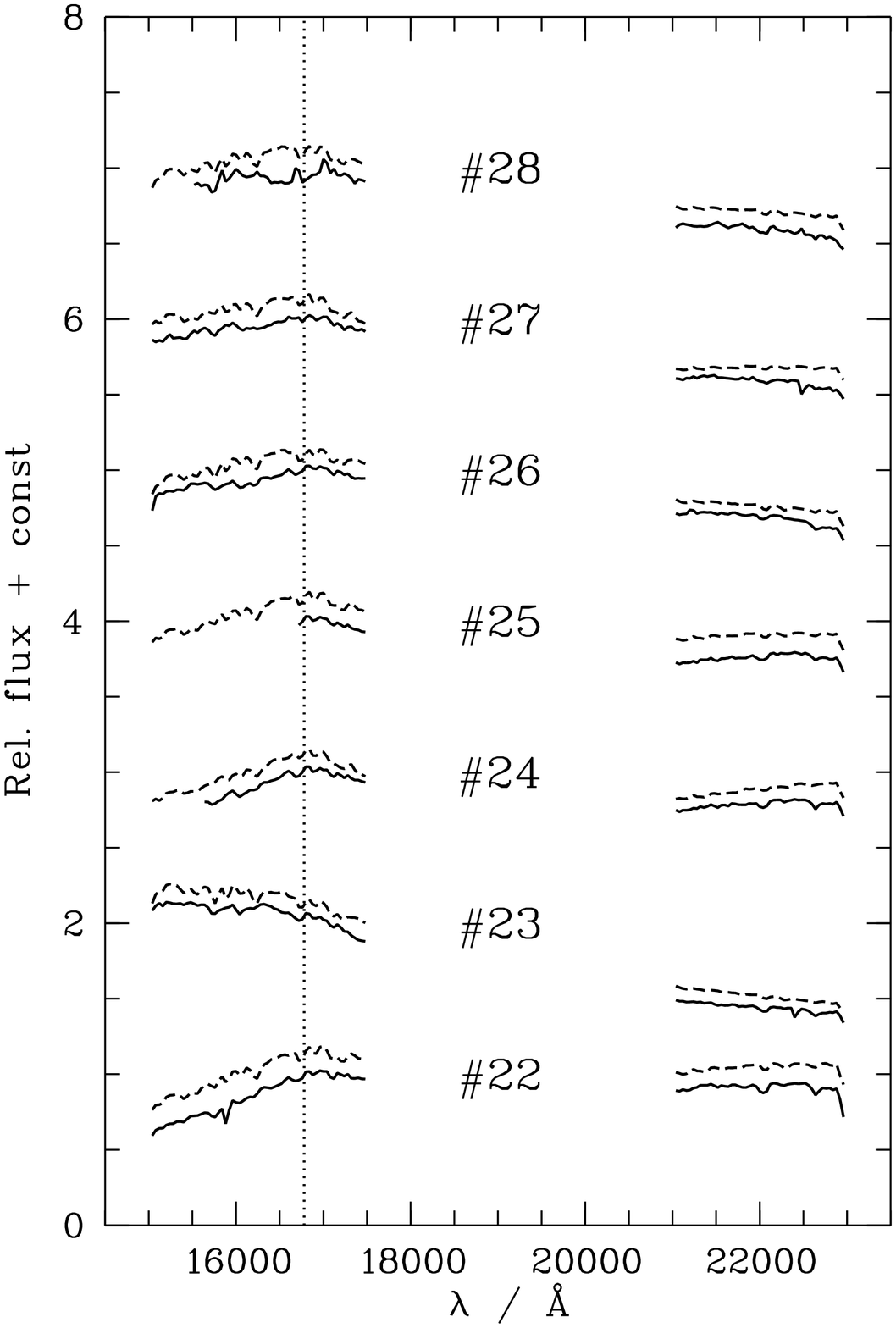}
\caption{Spectra of late-type objects in NGC~1333 (solid lines) in comparison with models (dashed lines), second part. The running number is the same as in Tables \ref{t1} and \ref{t2}. Shown is the best fit model with the effective temperature as given in the tables. Objects 1-19 are probable substellar members of NGC~1333.
\label{f5b}}
\end{figure*}

The low resolution of the spectra does not allow us to measure narrow atomic features. Instead, we focus on the broadband appearance, i.e.\ the shape of the H- and K-band as well as the relative intensity in these two bands. We benefit from the fact that the overall shape of the near-infrared spectrum changes distinctly at early/mid M spectral types. While earlier type stars show a broadband spectrum which is monotonously decreasing towards longer wavelength, this is not the case for $>$M5 objects, due to the appearance of strong H$_2$O absorption features at 1.3-1.5$\,\mu m$ and 1.75-2.05$\,\mu m$. Most notably, this causes the H-band spectrum to have a distinct and broad peak centered on 1.68$\,\mu m$, in contrast to earlier type objects, which increases in strength towards L-type objects. This fundamental change is discussed in detail in the literature, see for example \citet{2005ApJ...623.1115C}. As demonstrated in Fig.\ \ref{f4}, the H-band peak is dominant for temperatures lower than 3000\,K. While the strength and shape of the H-band peak is primarily determined by temperature (and gravity, see \ref{s42a}), there is an additional effect of extinction -- high extinction (see Sect. \ref{s32}), will make the H-band peak slightly sharper. As can be seen in Fig.\ \ref{f4}, this is clearly a minor effect and is only apparent when comparing over a wide range of extinctions.

\begin{figure*}[t]
\center
\includegraphics[width=5.9cm]{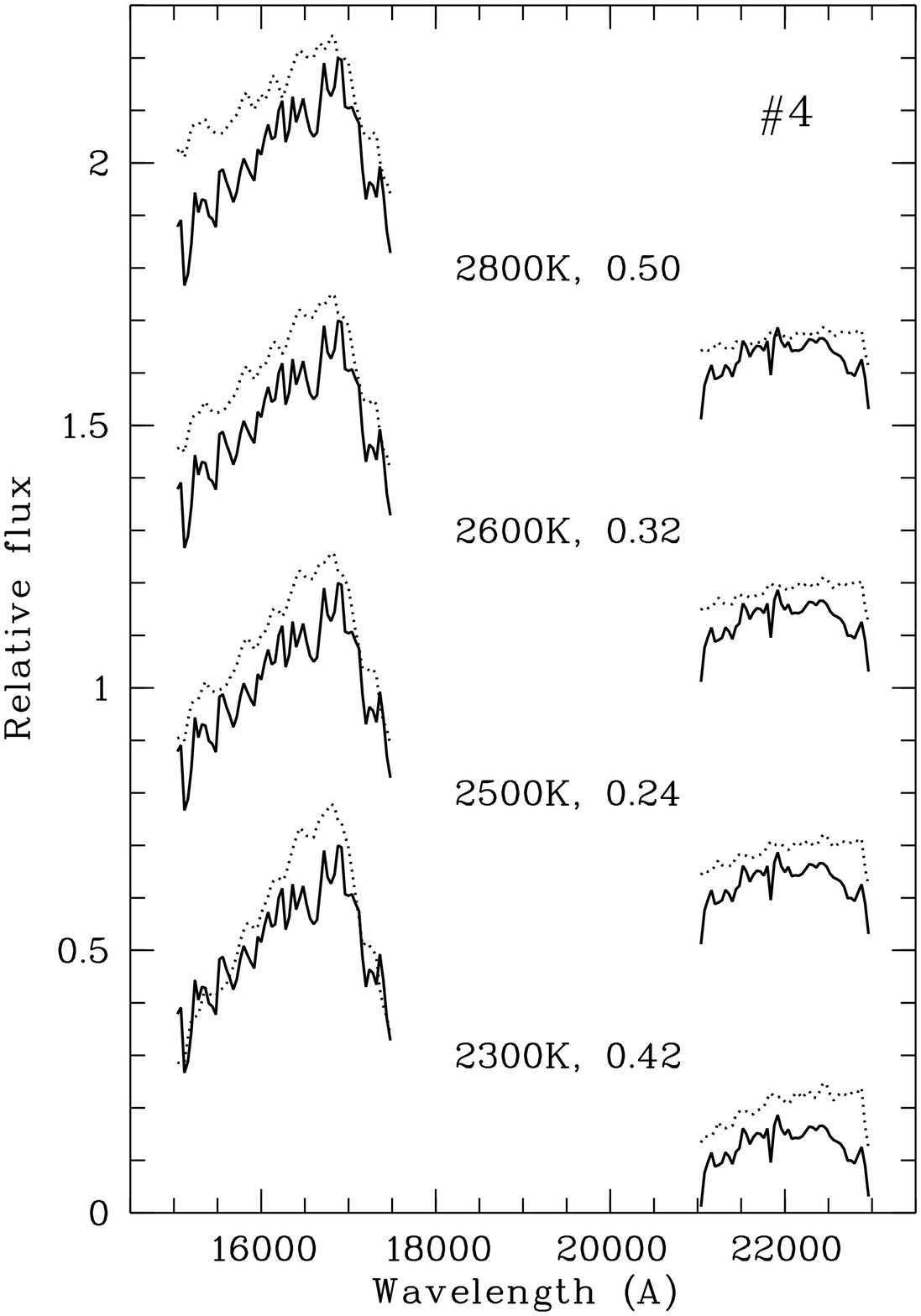} \hfill
\includegraphics[width=5.9cm]{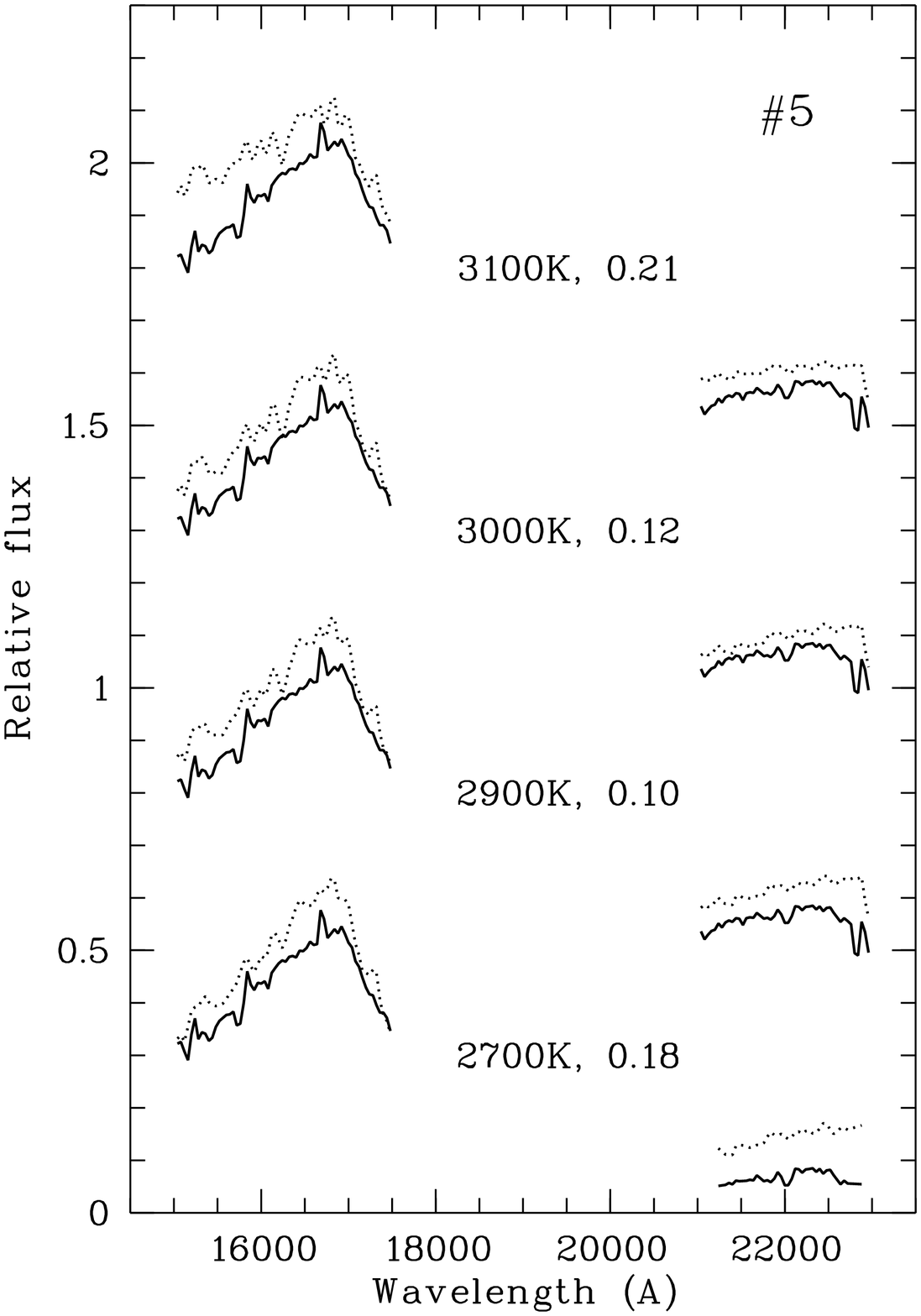} \hfill
\includegraphics[width=5.9cm]{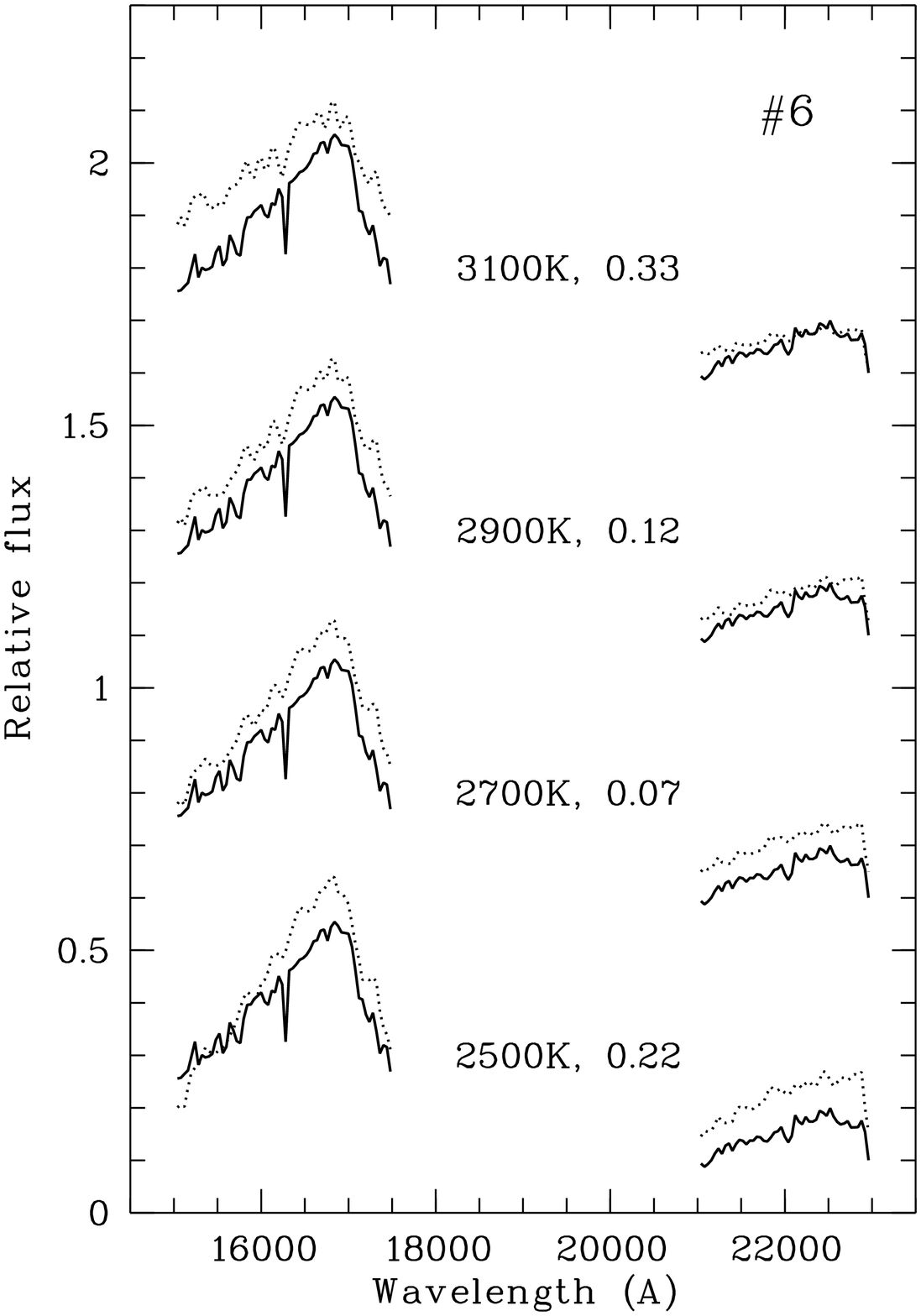}
\caption{Illustration of the spectral fitting process: Each panel shows the spectrum of a probable brown dwarf in NGC~1333 from Table \ref{t1} in comparison with model spectra for a range of effective temperatures. For clarity, the models are offset by 0.05 units. For each comparison, we give the effective temperature of the model and the $\chi^2$ of the fit. From the $\chi^2$ values we infer a conservative uncertainty of $\pm 200$\,K for the effective temperature. The plot demonstrates that larger offsets in $T_{\mathrm{eff}}$ lead to visible discrepancies between model and observation, particularly in the shape of the H-band peak. Note: For object \#4, the photometrically determined extinction of $A_V=2.3$\,mag gives $\chi^2 \ge 0.37$; much better results are achieved with a slightly lower $A_V$. We use here $A_V = 1$\,mag.
\label{f11}}
\end{figure*}

All spectra were examined visually to look for the typical signature of water absorption, mainly the characteristic peak in the H-band. The sample clearly falls into two groups, where 28/53 objects (53\%) show an indication of a peak while the others do not. In this classification, we erred on the side of caution, and therefore also included objects with only tentative evidence of water absorption. This sample of 28 objects is the `confirmed' sample of Figs.\ \ref{f2}-\ref{f7} and is listed in Tables \ref{t1} and \ref{t2}, together with the photometric and spectroscopic properties. All objects without a clear H-band peak are most likely stars with spectral types significantly earlier than M5. Our spectra do not permit definitive classification of these sources. Out of the 28 confirmed objects, 21 come from our photometric candidate sample. All nine objects from \citet{2004AJ....127.1131W} with spectral types M2-M8 are correctly classified as having an H-band peak. The `peak' objects exhibit other spectral features of late-type sources, most notably the CO bandheads at $>2.29\,\mu m$. We do not find evidence for CH$_4$ absorption at 1.6-1.8$\,\mu m$, typical for late L and T dwarfs.

We compared the 28 spectra with the H-band `peak' with the latest version of the spectra from the AMES DUSTY model atmospheres \citep{2001ApJ...556..357A}, which are based on the PHOENIX stellar atmosphere code. In contrast to the NEXTGEN spectra from the same code, DUSTY spectra include dust formation in the convective envelope and dust opacities. The more recent DUSTY models also include more complete H$_2$O and TiO line lists. To probe if the specific treatment of dust is relevant, we compared the DUSTY spectra with the DRIFT PHOENIX series \citep{2008A&A...491..311S}, which include a dust model and consistently treat cloud formation and its feedback on the atmosphere. For the resolution and wavelength coverage comparable with our observed spectra, there is no significant difference in the temperature regime $>2200$\,K. For lower temperatures, the dust treatment becomes critical and alternative models need to be considered \citep{2008ApJ...675L.105H}.

We used the DUSTY model spectra with temperatures of 1800-3900\,K in steps of 100\,K and the lowest available $\log{g}$ (3.5 or 4.0). The model spectra were binned and smoothed in the same way as the science spectra. All spectra are scaled to a flux of 1.0 at 1.7$\,\mu m$. Each object spectrum was then compared with the model series, reddened with the optical extinction derived in Sect.\ \ref{s32}. The quality of the match was judged by calculating the $\chi^2$ aided by visual inspection. To avoid being misled by excess continuum due to disks or accretion and uncertainties in the extinction, we put emphasis on matching the H-band shape, rather than the overall spectrum. 

In Figs.\ \ref{f5a} and \ref{f5b} we show all 28 object spectra with H-band peak (solid lines) and their best fit model spectrum (dashed lines). The ID numbers are the same as in Table \ref{t1} and \ref{t2}. The best match temperatures are given in the tables. For 18 out of 28 objects, this procedure yields a minimum $\chi^2$ of $<0.5$ and a well-matched spectrum, without further adjustments. Four more objects (\#4, \#10, \#20, \#24) fulfill the same criteria after minor changes to the extinction value ($\Delta A_V\le 2$\,mag). Object \#26 requires increasing the extinction from $A_V=4.4$ to 8\,mag, to obtain a good match, which results in $T=3800$\,K. As argued in Sect.\ \ref{s42}, the status of this object is unclear; it may be a red giant in the background of the cluster. Object \#12 shows steeply increasing flux levels in the K-band and all signs of strong reddening. We obtain a good match between model and observed spectrum by increasing the extinction to $A_V=18$\,mag. The source is only partially covered in the H-band, therefore the temperature estimate is preliminary. Based on the slope of the H-band spectrum, the best estimate is $T=2200-2800$\,K. Two objects (\#8, \#18) agree well with the model in the H-band, but show continuous excess emission in K-band with respect to the best fitting model. Both sources are found to have mid-infrared color excess due to the presence of circumstellar material (Sect.\ \ref{s44}), which could explain the K-band excess. 

Thus, for 26 out of 28 objects we obtain a decent match between the model and observed spectra. In Fig.\ \ref{f11} the typical uncertainty of the fitting procedure is illustrated by comparing the observed spectra for three objects from Table \ref{t1} with models for various effective temperatures. The $\chi^2$ values used to select the best-fitting model are listed as well. As can be seen in the figure, the procedure often does not allow us to distinguish reliably between two models with $\Delta T = 100$\,K. However, for $\Delta T \ga 200$\,K there are discernible differences, particularly in the H-band peak, between model and observation, combined with a clear increase in $\chi^2$. Thus, a conservative estimate for the typical uncertainty in $T_{\mathrm{eff}}$ is $\pm 200$\,K. For all $>$M5 objects, the spectral types -- either from \citet{2004AJ....127.1131W} or from the spectral index, see below -- are consistent with the temperatures from model fitting within $\pm 200$\,K or $\pm 1$ subtype (using the temperature scale from \citet{2003ApJ...593.1093L}). Temperatures of $>3500$\,K have to be treated with caution, as our model grid is limited at 3900\,K.

The two objects with unsatisfactory results in the fit are \#11 and \#15. For \#15, the H-band spectrum is well-matched with a temperature of 3000\,K which is consistent with the spectral type: M6.5 from \citet{2004AJ....127.1131W} and M7.5 from our spectrum, see below. The literature K-band spectrum shows a flux increase from 2.1 to 2.3$\,\mu m$, as expected for a mid/late M-type object. Thus, we consider the temperature estimate of $\sim 3000$\,K to be reliable. Our K-band spectrum declines with wavelength, indicative of a significantly higher temperature, which prevents us from obtaining a good fit. The object also has a mid-infrared color excess, indicative of circumstellar material (Sect.\ \ref{s44}). Source \#11 exhibits strong emission at 2.12$\,\mu m$, consistent with the 1-0 S(1) line of molecular hydrogen at 2.122$\,\mu m$, seen in protostellar outflows. At $\lambda >2.25\,\mu m$, the flux decreases sharply, which is not reproduced by any models or literature spectra in this temperature domain. Based on the H-band peak, our best estimate for the effective temperature is 2600\,K, but this has to be treated with caution. 

As a complementary approach, we computed the spectral index proposed by \citet{2007ApJ...657..511A} for spectral classification of young BDs, applicable from M5 to L5. This index has two distinct advantages compared with the variety of other index schemes in the literature, as it is gravity insensitive and does not require wavelength coverage in regions heavily affected by telluric water absorption. The spectral types from this index confirm the \citet{2004AJ....127.1131W} spectral types for the seven $>$M5 objects within $\pm 1.1$ subtypes. Where applicable, we use this index as an additional indicator for the spectral type of our objects (see Table \ref{t1}).

\subsection{Summary of substellar member selection}
\label{s34}

In summary, from our 36 photometrically selected candidates, we confirm 21 as M-type objects. In addition, three out of nine bluer objects and one out of five random objects are classified as M-type sources. Nine cluster members from \citet{2004AJ....127.1131W} are confirmed by our spectroscopy, seven of them overlapping with our photometrically selected sample. In total, we have 28 confirmed objects with a H-band peak, indicative of spectral type M. In this confirmed sample are 19 objects with spectral types $>$M5 or effective temperatures of $\le$3000\,K, which would qualify them as brown dwarfs, if they are in fact young members of NGC~1333 (see Sect.\ \ref{s42a}). This subsample is summarized in Table \ref{t1}, the remaining nine objects in Table \ref{t2}. For completeness, we note that four photometric candidates reported by \citet{2004AJ....127.1131W} do not show the H-band peak, i.e.\ are not classified as M-type sources: MBO 39, 59, 76, 109. These objects may still be cluster members with earlier spectral types.

\begin{deluxetable*}{cllccccccll}
\tabletypesize{\scriptsize}
\tablecaption{Probable substellar members of NGC~1333.\label{t1}}
\tablewidth{0pt}
\tablehead{
\colhead{no.} & \colhead{$\alpha$(J2000)} & \colhead{$\delta$(J2000)} & 
\colhead{i' (mag)}  & \colhead{z' (mag)} & \colhead{J (mag)} & \colhead{K (mag)} &
\colhead {$A_V$} & \colhead{$T_{\mathrm{eff}}$\tablenotemark{a}} & \colhead{SpT\tablenotemark{b}} & 
\colhead{Notes\tablenotemark{c}}}
\tablecolumns{11}
\startdata
 1 & 03 28 47.66 &+31 21 54.6 & 23.540 & 21.343 & 17.55 & 15.24 & 6.9  & 2800  &      & MBO139  		     \\ 
 2 & 03 28 54.93 &+31 15 29.1 & 21.040 & 18.922 & 15.99 & 14.22 & 4.1  & 2600  & M6.5 & ex\tablenotemark{d,e}	     \\ 
 3 & 03 28 55.25 &+31 17 35.4 & 19.386 & 17.602 & 15.09 & 13.43 & 3.5  & 2900  &      & ASR38\tablenotemark{d}       \\ 
 4 & 03 28 56.50 &+31 16 03.1 & 22.999 & 21.233 & 18.17 & 16.73 & 2.3  & 2500  &      & \tablenotemark{e}	     \\ 
 5 & 03 28 56.94 &+31 20 48.6 & 19.417 & 17.935 & 15.48 & 13.92 & 3.0  & 2900  & M6   & MBO91, ex		     \\ 
 6 & 03 28 57.11 &+31 19 12.0 & 21.157 & 19.545 & 17.24 & 15.34 & 4.8  & 2700  & M8   & MBO148, ex		     \\ 
 7 & 03 28 58.43 &+31 22 56.7 & 19.230 & 17.856 & 15.21 & 13.54 & 3.5  & 2800  & M6.5 & MBO80, ex		     \\ 
 8 & 03 29 03.39 &+31 18 39.9 & 20.059 & 18.199 & 15.85 & 14.03 & 4.3  & 2600  & M8.5 & MBO88, ex		     \\ 
 9 & 03 29 05.56 &+31 10 13.7 & 21.288 & 19.425 & 16.91 & 15.76 & 0.8  & 2600  & M8   & 			     \\ 
10 & 03 29 05.64 &+31 20 10.7 & 20.080 & 18.842 & 17.11 & 15.49 & 3.3  & 2500  &      & MBO143, ex\tablenotemark{d,e}\\ 
11 & 03 29 07.17 &+31 23 22.9 & 22.989 & 21.187 & 17.87 & 15.62 & 6.6  & (2600)&      & MBO141\tablenotemark{e}      \\ 
12 & 03 29 09.33 &+31 21 04.1 & 22.553 & 20.336 & 16.33 & 13.29 & 10.7 & (2500)&      & MBO70, ex\tablenotemark{e}   \\ 
13 & 03 29 10.79 &+31 22 30.1 & 19.483 & 17.778 & 14.96 & 13.05 & 4.8  & 3000  & M7.5 & MBO62			     \\ 
14 & 03 29 14.43 &+31 22 36.2 & 18.104 & 16.960 & 14.55 & 13.02 & 2.8  & 2900  & M7   & MBO66			     \\ 
15 & 03 29 17.75 &+31 19 48.1 & 19.091 & 17.326 & 14.80 & 12.99 & 4.3  & 3000  & M7.5 & MBO64, ex\tablenotemark{d,e} \\ 
16 & 03 29 28.16 &+31 16 28.4 & 16.934 & 15.808 & 13.22 & 12.53 & 0.0  & 2600  & M7.5 & random  		     \\ 
17 & 03 29 33.88 &+31 20 36.1 & 20.383 & 18.848 & 16.47 & 15.37 & 0.5  & 2500  & M8   & MBO140  		     \\ 
18 & 03 29 35.71 &+31 21 08.5 & 22.633 & 21.089 & 18.50 & 16.94 & 3.0  & 2500  &      & blue, ex		     \\ 
19 & 03 29 36.36 &+31 17 49.8 & 22.528 & 20.716 & 17.91 & 16.38 & 2.8  & 2700  &      & 			     \\ 
\enddata
\tablenotetext{a}{Estimated by comparing the spectra to models. Uncertain fit results are in brackets.}
\tablenotetext{b}{Estimated using the spectral index and calibration from \citet{2007ApJ...657..511A}, if applicable.}
\tablenotetext{c}{`ex' means mid-infrared color excess, ASR and MBO object id's are candidates from \citet{2004AJ....127.1131W}.}
\tablenotetext{d}{Near-infrared photometry from 2MASS.}
\tablenotetext{e}{Temperature estimate changed after visual inspection, see Sect.\ \ref{s33}.}
\end{deluxetable*}

\begin{deluxetable*}{cllccccccll}
\tabletypesize{\scriptsize}
\tablecaption{Possible stellar members of NGC~1333.\label{t2}}
\tablewidth{0pt}
\tablehead{
\colhead{no.} & \colhead{$\alpha$(J2000)} & \colhead{$\delta$(J2000)} & 
\colhead{i' (mag)}  & \colhead{z' (mag)} & \colhead{J (mag)} & \colhead{K (mag)} &
\colhead {$A_V$} & \colhead{$T_{\mathrm{eff}}$\tablenotemark{a}} & \colhead{SpT\tablenotemark{b}} 
& \colhead{Notes\tablenotemark{c}}}
\tablecolumns{11}
\startdata
20 & 03 28 43.23 &  +31 10 42.6 & 21.864 & 19.606 & 15.37 & 12.03 & 12.3 & 3400 & & ex\tablenotemark{d,e}  	        \\ 
21 & 03 28 47.34 &  +31 11 29.9 & 21.626 & 19.330 & 15.48 & 12.70 & 9.4  & 3100 & & \tablenotemark{d}		        \\ 
22 & 03 28 50.97 &  +31 23 47.9 & 20.380 & 18.418 & 14.66 & 11.24 & 12.7 & 3500 & & MBO26\tablenotemark{d} 	        \\ 
23 & 03 29 02.16 &  +31 16 11.1 & 17.197 & 16.245 & 14.53 & 13.57 & 0.0  & 3900 & & ASR03, giant?\tablenotemark{d}      \\ 
24 & 03 29 08.32 &  +31 20 20.3 & 19.376 & 17.809 & 14.80 & 12.48 & 7.0  & 3100 & & MBO44\tablenotemark{e} 	        \\ 
25 & 03 29 11.64 &  +31 20 37.5 & 20.544 & 18.738 & 15.43 & 12.67 & 9.3  & 3400 & & MBO58\tablenotemark{d} 	        \\ 
26 & 03 29 16.75 &  +31 23 25.3 &        &        & 15.42 & 13.58 & 4.4  & 3800 & & MBO79, ex, giant?\tablenotemark{d,e}\\ 
27 & 03 29 22.36 &  +31 21 36.8 & 21.407 & 20.197 & 17.71 & 15.89 & 4.3  & 3300 & & MBO177, blue 			\\ 
28 & 03 29 33.50 &  +31 17 56.5 & 22.926 & 21.477 & 18.52 & 16.27 & 6.6  & 3700 & & blue				\\ 
\enddata
\tablenotetext{a}{Estimated by comparing the spectra to models. Uncertain fit results are in brackets.}
\tablenotetext{b}{Estimated using the spectral index and calibration from \citet{2007ApJ...657..511A}, if applicable.}
\tablenotetext{c}{`ex' means mid-infrared color excess, ASR and MBO object id's are candidates from \citet{2004AJ....127.1131W}.}
\tablenotetext{d}{Near-infrared photometry from 2MASS.}
\tablenotetext{e}{Temperature estimate changed after visual inspection, see Sect.\ \ref{s33}.}
\end{deluxetable*}

\section{Characterization of confirmed late-type objects}
\label{s4}

\subsection{Effective temperatures}
\label{s42}

We have determined effective temperatures for the sample of late-type objects in Tables \ref{t1} and \ref{t2} by fitting the observed to model spectra. Out of 28 sources, 19 have temperatures between 2500 and 3000\,K, for the other nine we obtain higher values. In addition, we have already derived extinctions, based on the $J-K$ colors, and dereddened J-band magnitudes (see Sect.\ \ref{s32}, Fig.\ \ref{f7}). 

We use the dereddened J-band photometry to obtain an alternative estimate of effective temperatures by comparing with the BCAH98 evolutionary tracks BCAH98 \citep{1998A&A...337..403B} for the typical cluster age of 1\,Myr. This relies on the assumption that the objects are in fact young members of the cluster NGC~1333, for which we adopt a distance of 300\,pc, consistent with the estimates by \citet{1999AJ....117..354D,2002A&A...387..117B}. The J-band is the wavelength regime that is normally assumed to be least affected by accretion, disk excess, and magnetic activity. It is found that the BCAH98 temperature vs. J-mag relation is well approximated by a fourth degree polynomial. Applying this function to the dereddened J-band photometry yields `photometric' temperatures from 2100 to 3400\,K. 

For 15 objects, the photometric and spectroscopic temperatures agree within $\pm 200$\,K, confirming our assessment of $T_{\mathrm{eff}}$ given in Sect.\ \ref{s33}. Four objects (23, 26-28 in Table \ref{t2}) have photometric temperatures 800-1200\,K lower than the ones from the spectroscopy, which can be explained with being background objects. Assuming intrinsic colors for dwarf stars from the 1\,Gyr isochrone in the BCAH98 model tracks puts their distances at 360-950\,pc. The remaining objects have offsets of 200-500\,K between the two values, and in most cases the photometric temperatures are too low. One possible explanation is an age spread in the cluster. As shown by \citet{2009AJ....137.4777W}, the members of NGC~1333 are significantly dispersed in the HR-diagram, possible indicating an age spread from $<1$\,Myr to several Myrs. Furthermore, the model tracks have well-documented uncertainties in this mass and age regime \citep{2002A&A...382..563B}.

For two objects, \#23 and \#26, there is a clear discrepancy between our spectroscopic temperatures of 3800 and 3900\,K and the spectral types of M2.6 and M4.2 derived by  \citet{2004AJ....127.1131W}, which would imply temperatures of only 3500 and 3300\,K, respectively. As pointed out above, these are likely to be background objects. The K-band index for spectral typing used by \citet{2004AJ....127.1131W} gives significantly earlier spectral types for giants than for dwarfs (about M0 or earlier for these two objects). These types would be consistent with our effective temperatures. Thus, these two objects could be red giants in the background of NGC~1333. However, assuming the intrinsic giant colors from \citet{1988PASP..100.1134B}, this would put them as an unreasonably large distance of 30\,kpc for spectral type M0. A significantly earlier spectral type is at odds with our spectroscopy. Thus, the nature of these objects remains uncertain.

\subsection{Cluster membership}
\label{s42a}

For the likely brown dwarfs in our sample (see Table \ref{t1}), the best evidence for youth and thus cluster membership is given by the shape of the H-band peak in the spectra. For substellar objects, the H-band peak appears to be gravity sensitive. Low gravity and thus youth is signified by a distinctly sharp peak, as opposed to the more rounded peaks seen in old objects \citep{2006ApJ...639.1120K}. The same characteristic is also seen in the available model spectra \citep{2001ApJ...556..357A}. 

\begin{figure*}
\center
\includegraphics[width=8.8cm]{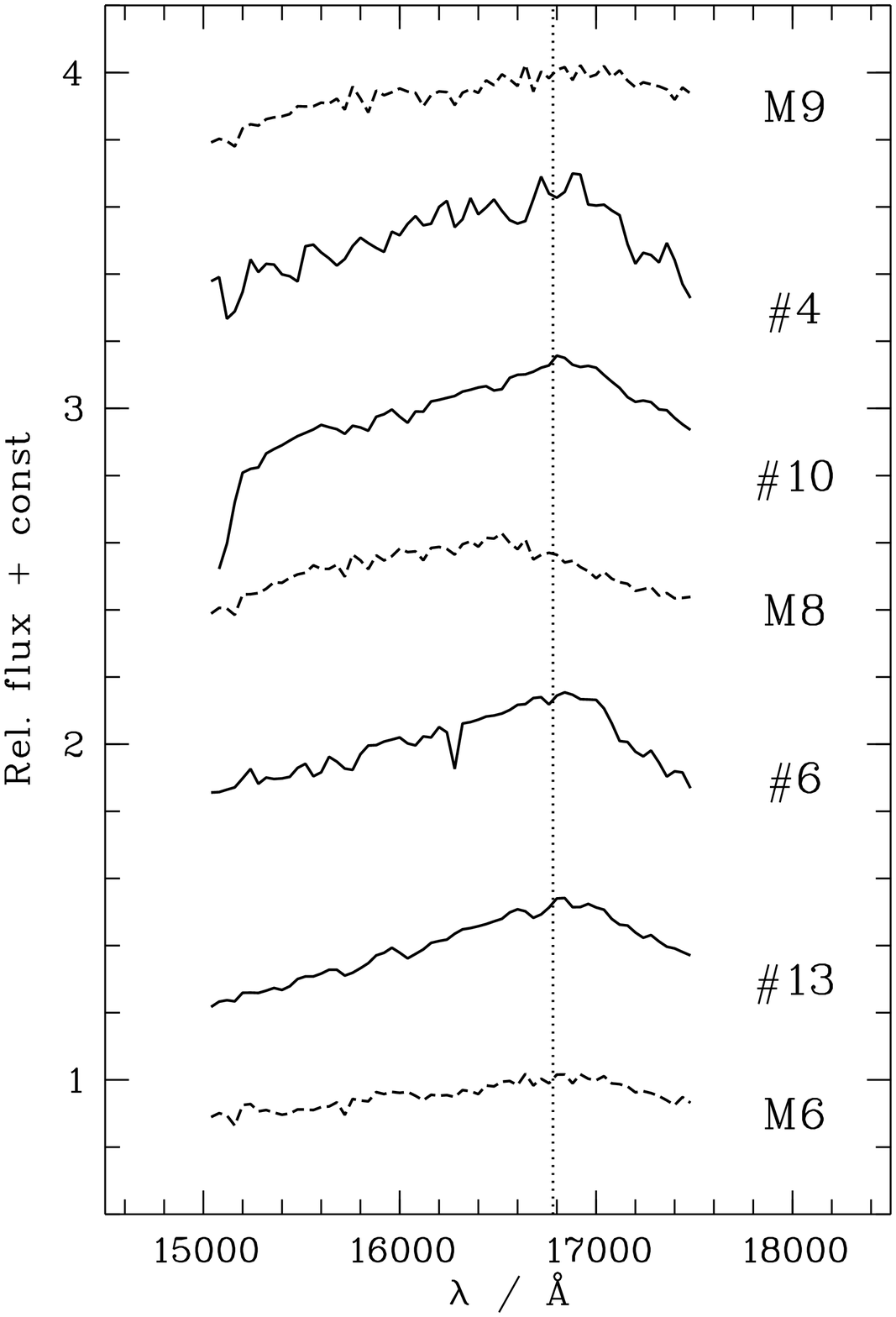} 
\includegraphics[width=8.8cm]{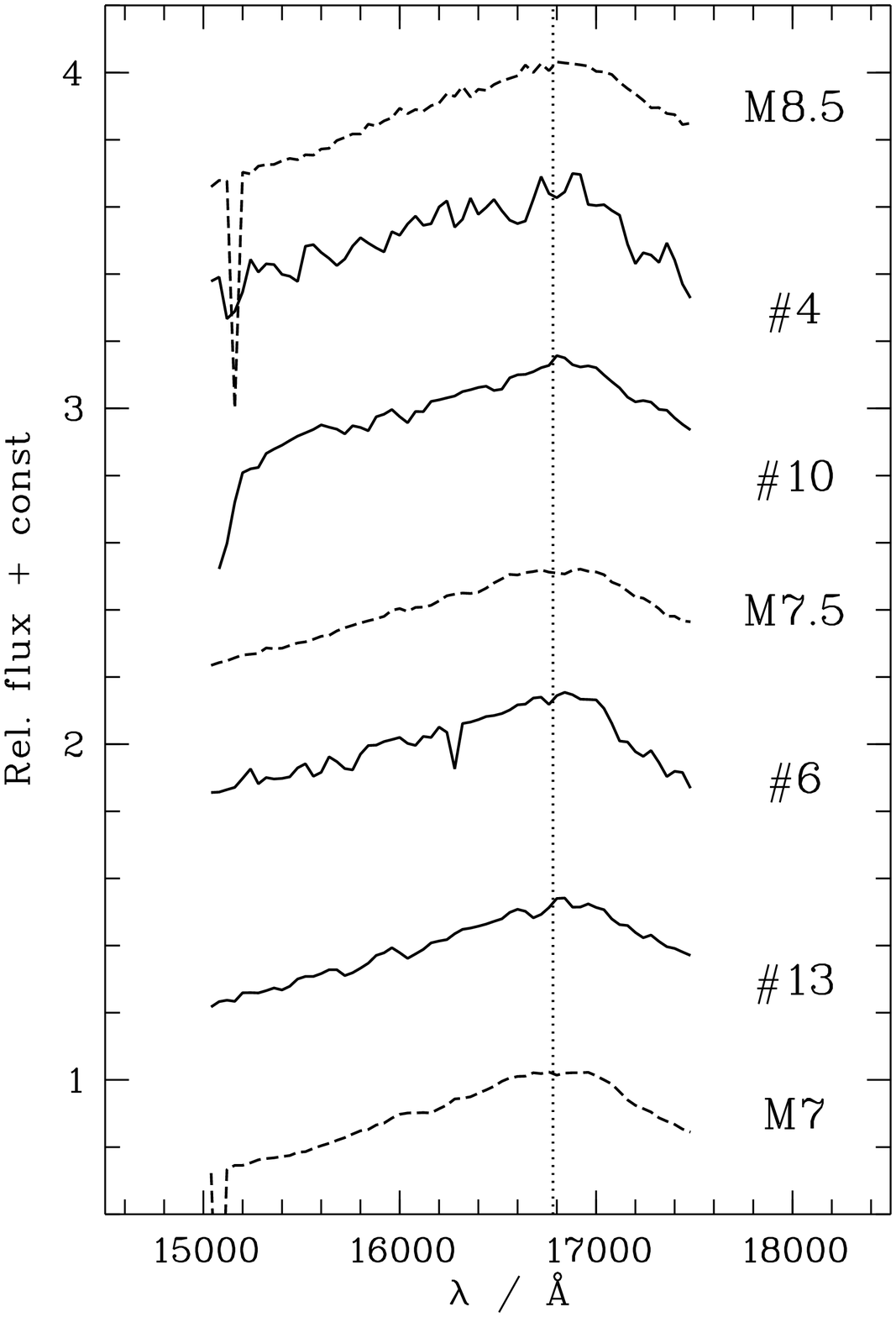} 
\caption{Spectra of very low-mass objects in NGC~1333 from Table \ref{t1} (solid) in comparison with spectra of field dwarfs from the NIRSPEC Brown Dwarf Spectroscopic Survey \citep[left panel,][]{2003ApJ...596..561M} and with young brown dwarfs in Taurus \citep[right panel,][]{2007AJ....134..411M}. The field dwarfs are from bottom to top Wolf\,359 (M6), VB\,10 (M8), LHS\,2065 (M9); the Taurus BDs are from bottom to top MHO\,4 (M7), KPNO\,1 (M7.5), and KPNO\,5 (M8.5). The objects in NGC~1333 have a significantly `sharper' peak than the field dwarfs, but are well matched by the young Taurus BDs. To account for the higher extinctions in NGC~1333, the literature spectra have been reddened to $A_V=5$\,mag.
\label{f6}}
\end{figure*}

All 19 sources with effective temperatures of 3000\,K or lower listed in Table \ref{t1} show a sharp peak. As an example, we show four of them in comparison with published spectra for field dwarfs of similar spectral type from the NIRSPEC Brown Dwarf Spectroscopic Survey \citep{2003ApJ...596..561M} in Fig. \ref{f6}, left panel. The difference in the shape of the H-band feature is obvious. Particularly, the slope on the red side of the peak is significantly steeper in our objects. On the other hand, our spectra agree well with those of young brown dwarfs in Taurus, taken from \citet{2007AJ....134..411M}, as shown in Fig. \ref{f6}, right panel.

Statistically, we do not expect a significant contamination with late field M dwarfs in our sample in Table \ref{t1}. According to the data compiled in \citet{2008A&A...488..181C}, the space density of M6-7 dwarfs is $4.9 \cdot 10^{-3}$\,pc$^{-1}$ and drops steeply to $1.3\cdot 10^{-3}$\,pc$^{-1}$ for M8-9. Our sample covers an area of about 200\,arcmin$^2$ which corresponds to $\sim 1.5$\,pc$^2$ at the distance of NGC~1333. In a cone with this footprint and 300\,pc height we expect about one M6-M9 field dwarf. Thus, for the M6 or later objects in Table \ref{t1} the contamination by field objects is negligible.

We conclude that the $>$M5 objects identified in this survey (`probable' members in Table \ref{t1}) are in fact young substellar members of the cluster NGC~1333. For the earlier type stars (`possible members' in Table \ref{t2}), significant contamination by field dwarfs or giants is more likely, as their space density is larger and the H-band signature of youth is less pronounced. Further evidence for youth comes from the fact that many of our objects show mid-infrared color excess indicative of the presence of circumstellar material, as discussed in Sect.\ \ref{s44}, and from the spatial clustering around the main cloud core in NGC~1333, see Sect.\ \ref{s43}.

\subsection{Masses}
\label{s42b}

We estimate the masses of the probable brown dwarfs in our sample by comparing the effective temperatures and the dereddened J-band photometry with the theoretical model tracks for an assumed age. Considering the uncertainties in model tracks and object ages, we refrain from deriving masses for individual objects. Instead, we focus on establishing the approximate mass limits in our sample. 

The range of effective temperatures in this sample is 2500-3000\,K. Comparing directly to the 1-3\,Myr tracks from BCAH98 yields masses of 0.02-0.1$\,M_{\odot}$. The result is very similar for the COND03 and DUSTY00 tracks. If some objects are younger than 1\,Myr, they will be slightly more massive for a given temperature. Taking into account 200\,K uncertainty in our temperature estimate, the lowest mass limit becomes 0.012-0.015$\,M_{\odot}$, according to the COND03 and DUSTY00 tracks. Based on the dereddened J-band photometry, the mass range in our brown dwarf sample is 0.006-0.1$\,M_{\odot}$ for the 1\,Myr isochrones, or 0.012-0.2$\,M_{\odot}$ for the 5\,Myr isochrones. Note that the latter values are not significantly affected by uncertainties in the distance to NGC~1333. The most recent distance estimates agree fairly well at $\sim 300$\,pc with an uncertainty of about 30\,pc \citep{1999AJ....117..354D,2002A&A...387..117B}, corresponding to $\pm 0.2$\,mag or about 10\% in object mass.

There are two reasons why we put more trust in the mass estimates obtained from the temperatures: 1) Luminosities depend much more on age than temperatures on the pre-main sequence, thus the mass estimates based on temperatures are more robust against age spread, which may be significant in this cluster \citep{2009AJ....137.4777W}. The Lyon evolutionary tracks are known to underestimate object masses derived from fluxes. For example, for the young brown dwarf eclipsing binary in ONC \citep{2006Natur.440..311S} the BCAH98 tracks predict a mass of 0.055-0.075$\,M_{\odot}$ based on its J-band magnitude (assuming a distance of $435\pm 55$\,pc); the actual system mass is 0.09$\,M_{\odot}$. More indirectly, \citet{2004ApJ...609..885M} find that the models predict brown dwarfs to be too bright in J-band at a given mass, by up to one order of magnitude. Both tests indicate that the masses from the model tracks, when inferred from luminosities, may systematically underestimate the actual masses by a factor of up to two. 

Based on the comparison of effective temperatures with the three model tracks, we conclude that the lowest mass limit in our brown dwarf sample is 0.012-0.02$\,M_{\odot}$. At the high mass end, the mass range of the objects in Table \ref{t1} extends across the substellar boundary into the stellar regime. Thus, a few of these objects may turn out to be very low-mass stars, instead of brown dwarfs.

\subsection{Spatial distribution and binarity}
\label{s43}

\begin{figure*}
\center
\includegraphics[width=16cm]{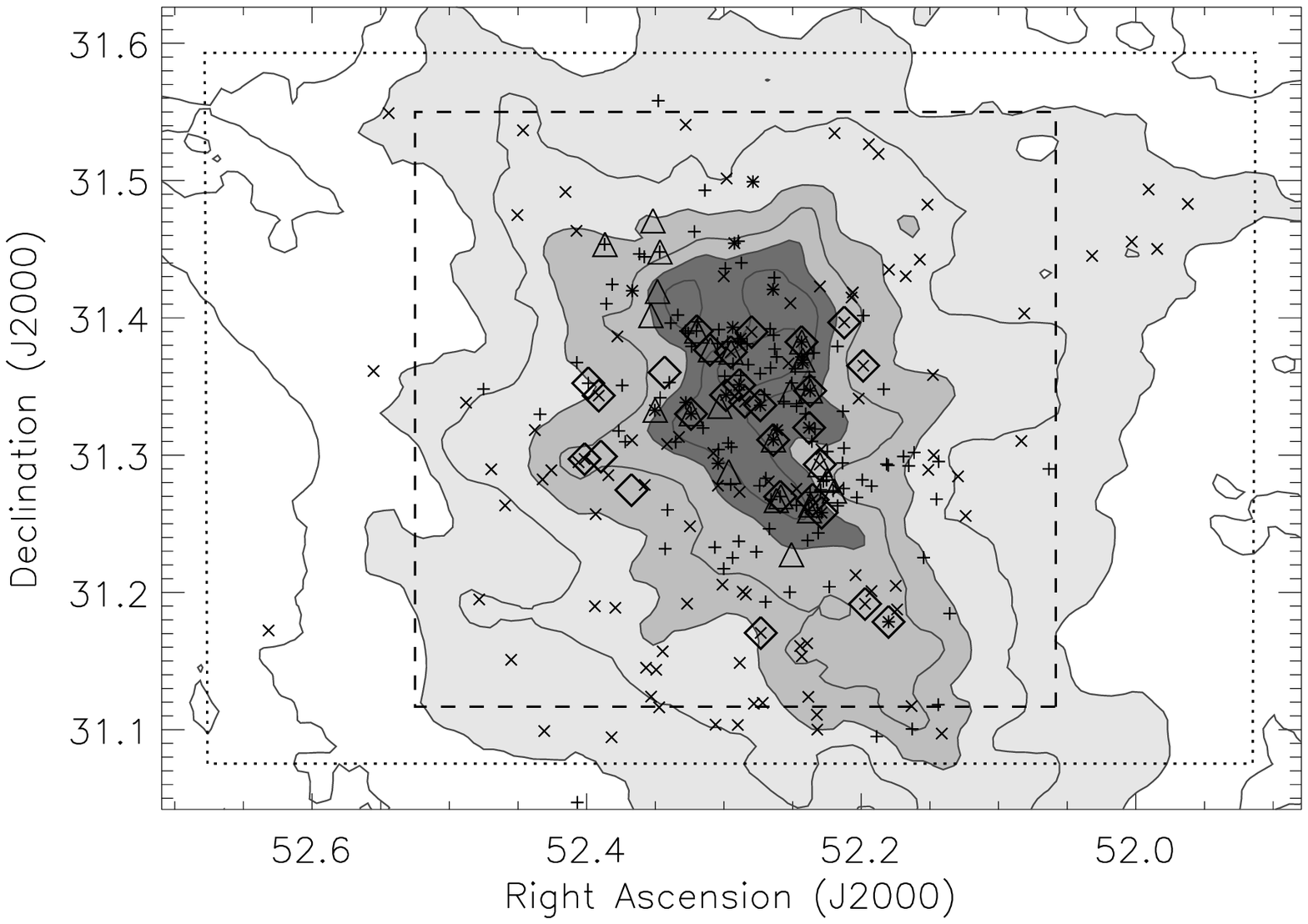} 
\caption{Spatial coverage of our survey and distribution of objects. Crosses are all candidates selected from optical photometry with 
$i' > 18.5$\,mag (Sect.\ \ref{s31}), diamonds the spectroscopically confirmed objects, triangles the confirmed brown dwarfs from \citet{2004AJ....127.1131W}. The plusses are YSO candidates selected from Spitzer data without spectroscopic confirmation from \citet{2008ApJ...674..336G}. The underlying contours show the distribution of $^{13}$CO from the $J=1-0$ map by \citet{2006AJ....131.2921R}. The dotted and dashed lines show the area covered in our optical and near-infrared survey, respectively.
\label{f9}}
\end{figure*}

In Fig.\ \ref{f9} we show the position of our photometric candidates and the confirmed members of NGC~1333 in relation to the gas in the cloud and other samples of cluster members. As already pointed out in Sect.\ \ref{s31}, the photometric candidates show a strong clustering towards the center of the cloud. On one side, this can be explained with the increasing extinction, causing background objects to appear heavily reddened. On the other hand, as seen in the plot, YSOs candidates selected from Spitzer mid-infrared excess \citep{2008ApJ...674..336G} show the same strong clustering, indicating that the distribution of members is very compact and does not extend to the edges of our survey field. Our spatial coverage thus fully encompasses the extent of the cluster. The small number of photometric candidates in the outer region is also evidence for the lack of contamination in our sample, as argued in Sect.\ \ref{s42a}.

All our spectroscopy fields were located within $\pm 0.1$\,arcmin of the cluster center, an area which contains the majority of known cluster members. As discussed in Sect.\ \ref{s33}, the `success rate' in finding late-type sources is 58\% (21/36) in the photometrically selected sample, indicating a substantial rate of contamination among the remaining photometric candidates. Due to our limited spatial coverage in the spectroscopy, we cannot give constraints on the nature of the (few) candidates outside the cloud core. Since the surface density of members drops rapidly with increasing distance from the center, we expect most of the outlying candidates to be foreground or background sources. 

The very compact distribution of brown dwarfs agrees with the more detailed analysis for the distribution of stars in NGC~1333 provided by \citet{2008ApJ...674..336G}, who report a steep decline of the density of members for cluster radii larger than 0.3\,pc (0.06\,degree at 300\,pc). \citet{2008ApJ...674..336G} also identify clear asymmetry in the distribution, attributed to the structure of the underlying cloud.  This all suggests that the cluster has experienced only limited dynamical evolution and most of the population in the core is in fact very young. Ages ranging from less than 1\,Myr to 3\,Myr have been inferred from HR diagram fitting for the objects concentrated in the molecular cloud gas \citep{2009AJ....137.4777W}. 

We visually checked the deep i'- and z'-band images for companions within a radius of 3.0\arcsec~of each of our spectroscopically confirmed M-type sources. The lower separation limit for the detection of companions is set by the seeing and the rejection criterion for elongated objects (see Sect. \ref{s21}); we determine this to be at $\sim$1.0\arcsec~(Sect.\ \ref{s21}), corresponding to a separation range of 300 to 1000\,AU for the distance of NGC~1333. No companions were found. For substellar objects in NGC~1333 (Table \ref{t1}) this gives a $2\sigma$ upper limit for binarity in the given separation range of 19\%. 

A number of recent studies clearly show a dearth of wide binaries in the very low-mass regime. In the central parts of the Cha~I star forming region, for example, a binary fraction of 11\% is found for separations $>20$\,AU \citep{2007ApJ...671.2074A}, and zero systems with separations larger than $50$\,AU. Our result in NGC~1333 is consistent with these findings \citep[for a review on VLM binarity see][]{2007prpl.conf..427B}. 

\subsection{Spitzer photometry}
\label{s44}

\begin{figure}
\center
\includegraphics[angle=-90,width=8.5cm]{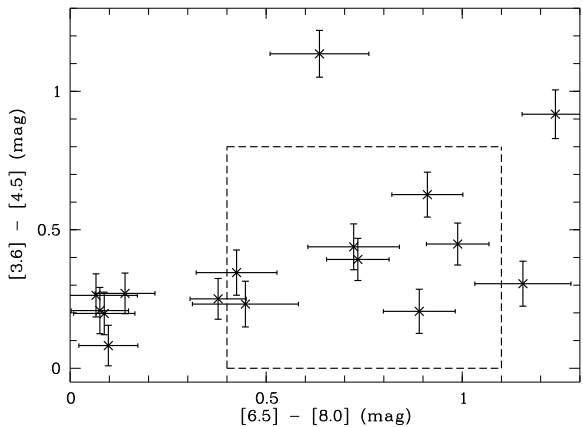} 
\caption{IRAC color-color diagram for the 16 objects from Tables \ref{t1} and \ref{t2} which have errorbars $<0.2$\,mag in all four bands. The approximate locus of the Class II objects is overplotted as a dashed square \citep{2004ApJS..154..363A}. Typical photospheric colors for late-type objects are around zero. Eleven objects have colors consistent with color excess due to circumstellar material.
\label{f8}}
\end{figure}

The NGC~1333 cluster is part of the Perseus star-forming complex and thus has been covered by the Spitzer Legacy Program `From Cores to Disks' (C2D, PI: Neil Evans). From the C2D source catalogues (version S13, full Perseus catalogue) we obtained the IRAC and MIPS fluxes from 3.6 to 24$\,\mu m$ for all objects in Tables \ref{t1} and \ref{t2}. Out of 28 sources, 16 are well-detected with errorbars $<0.2$\,mag in all four IRAC bands. The standard IRAC color-color diagram is shown in Fig.\ \ref{f8}. The color range for Class II objects is indicated. Eleven objects clearly show IRAC color excess, and thus are likely to be surrounded by circumstellar material. Ten of them are also safely detected at 24$\,\mu m$, with magnitudes $>2$\,mag larger than at 8$\,\mu m$. This confirms youth and thus membership to NGC~1333 for a large fraction of our sample. Objects with mid-infrared color excess are marked with `ex' in Table \ref{t1}. The disk fraction in our (small) sample is $\sim 70$\%, consistent with previous findings for stars \citep[83\%,][]{2008ApJ...674..336G} and brown dwarfs \citep[74\%,][]{2004AJ....127.1131W}.

The C2D data has recently been analysed in detail by \citet{2008ApJ...674..336G}. In this survey, all 11 objects with disks in our sample are classified as Class II sources. Two more objects in our sample (\#23 and \#25) are Class II objects with disks according to \citet{2008ApJ...674..336G}, but their IRAC errors are too large to be considered for Fig.\ \ref{f8}. 

\section{The mass function in NGC~1333}
\label{s5}

\subsection{Ratio of stellar to substellar members}
\label{s51}

Combining our survey with the ones by \citet{2004AJ....127.1131W} and \citet{2007AJ....133.1321G} there is now a substantial sample of spectroscopically confirmed brown dwarfs in NGC~1333. We use this to improve the constraints on the ratio of low-mass stars to brown dwarfs, a number recently measured for a range of other star-forming regions. We identify 19 objects as young M6 or later sources. Seven of them have previously been confirmed by \citet{2004AJ....127.1131W}. Their survey has found nine more objects with M6 or later spectral type, which we consider to be good candidates for brown dwarfs. The study by \citet{2007AJ....133.1321G} has limited spatial coverage (3\,arcmin$^2$), but finds five more objects in this spectral regime. All these objects are located close to the cluster center, which points to a high likelihood of being members. Thus, the current total census of likely brown dwarfs in this cluster is 33. As there are a number of additional photometric candidates in the cluster region (see Fig.\ \ref{f9}) that lack spectroscopic confirmation, this number should be treated as a lower limit; on the other hand, as pointed out in Sect.\ \ref{s42b}, some of these objects may be slightly above the substellar boundary.

To constrain the number of stellar members in NGC~1333, we use the Spitzer selected sample by \citet{2008ApJ...674..336G}, which identifies 137 likely YSOs in the cluster, 93 of them also have 2MASS near-infrared magnitudes. From these, 91 are in our survey area. We dereddened the photometry for these objects based on the $J-K$ color, using the procedure discussed in Sect.\ \ref{s32}. Comparing the dereddened J-band magnitude with the BCAH98 evolutionary track allows us to select objects in a given mass range. In this sample, 38 sources are likely to have masses between 0.08 and 1$\,M_{\odot}$. Based on the spectroscopic follow-up by \citet{2009AJ....137.4777W}, this sample is unlikely to have significant field star contamination. Spitzer only finds objects with disks; in fact, the cluster is known to harbour a significant population of disk-less T Tauri stars \citep{2009AJ....137.4777W}. Assuming a disk fraction of 83\% \citep{2008ApJ...674..336G} gives a total number of $\sim 50$ stellar cluster members in this mass regime. Taking into account the uncertainties in our mass cuts, this number may be inaccurate by up to 20\%.

Although the Spitzer sample of stars has been identified in a way not comparable to our own brown dwarf survey (and the two other ones in the literature), we are confident that it gives a robust estimate for the number of stars in this mass range in the cluster. The range of extinctions and the spatial distribution are similar in the Spitzer sample and the brown dwarf sample. Moreover, the disk fraction among the brown dwarfs is similar to the one seen in stars (Sect.\ \ref{s44}). Thus, in terms of evolutionary state and spatial coverage the stellar sample is representative.

From these numbers, we estimate the ratio of (low-mass) stars ($0.08<M<1.0\,M_{\odot}$) to brown dwarfs ($0.015<M<0.08\,M_{\odot}$) in NGC~1333 to be $50/33$, i.e.\ $1.5 \pm 0.3$. This can directly be compared with the literature. In an IMF analysis for seven different star-forming regions (not including NGC~1333), \citet{2008ApJ...683L.183A} find this ratio to be 3.3-8.5 depending on the region, with the lowest values measured in the ONC, NGC~2024, and Chamaeleon. Taking into account the 1$\sigma$ uncertainty in their values, the minimum ratio becomes 1.9 in Chamaeleon. These numbers are still significantly higher than the ratio in NGC~1333, pointing to an overabundance of brown dwarfs in NGC~1333 by a factor of 2-5.

This result is consistent with the previous findings for this cluster. \citet{2004AJ....127.1131W} derived a ratio of stars ($0.08<M<1.0\,M_{\odot}$) to very low-mass objects ($0.04<M<0.1\,M_{\odot}$) of $1.1 \pm ^{0.8}_{0.4}$, and considered this to be a lower limit, i.e.\ fully comparable with our result. (This would become 2.0, if they assume an age of 1\,Myr instead of 0.5\,Myr.) Including their lower mass brown dwarfs, \citet{2007AJ....133.1321G} find a value of $0.64\pm 0.38$ for this ratio. Although these numbers are not fully comparable with our estimate, due to the slightly different mass ranges, they are all clearly lower than the respective ratios in other regions. Thus, the mass function of NGC~1333 has a clear anomaly with an overabundance of brown dwarfs relative to 
low-mass stars. 

\subsection{The minimum mass in NGC~1333}
\label{s52}

The depth and large spatial coverage of our survey allows us to put meaningful limits on the abundance of planetary-mass objects in NGC~1333 in comparison with other regions. Recapitulating, the coolest objects in NGC~1333 have effective temperatures of 2500\,K, corresponding to masses of 0.012-0.02$\,M_{\odot}$. That means that there is no planetary-mass object in our spectroscopically confirmed sample. In the following, we will compare with the mass completeness limits in our survey and the frequency of planemos in other star-forming regions. Given the considerable uncertainties in the model evolutionary tracks in this age and mass regime, we will follow two guidelines: a) Conversion from effective temperatures to masses is more reliable than from luminosities (or magnitudes) to masses, see the discussion in Sect.\ \ref{s42b}. b) We prefer to base the arguments on relative comparisons consistently using the same model tracks, and do not place strong emphasis on the absolute mass values. 

Our coolest and lowest mass objects in the spectroscopic sample have $T_{\mathrm{eff}}=2500\pm 200$\,K, i.e.\ we rule out the presence of objects with 2200\,K or cooler in this group of objects. The five faintest confirmed brown dwarfs have 22.5-23.5\,mag in the i'-band with extinctions $A_V\la 10$\,mag. Our completeness limit is 24.7\,mag in i'-band, i.e.\ 1.2-2.2\,mag deeper than the faintest confirmed objects. Based on the 1\,Myr isochrones from COND03 and DUSTY00, an object that much fainter than one with 2500\,K has 0.006-0.01$\,M_{\odot}$, with $T_{\mathrm{eff}}$ of 2000-2200\,K, i.e. for $A_V\la 10$\,mag we are complete down to $\sim 0.008\,M_{\odot}$. For $A_V\la 5$\,mag the limit becomes $\sim 0.004\,M_{\odot}$ or 1800\,K. Thus, our survey is complete to masses significantly below the low-mass limit of the confirmed objects. In fact, six objects with i'-mag $>23.5$\,mag have been observed spectroscopically and turned out not to have an H-band peak, indicative of low temperatures. 

As discussed in Sect.\ \ref{s1}, previous surveys with comparable depth (in terms of masses) have been carried out in the $\sigma$\,Ori cluster, the Orion Nebula Cluster (ONC), and the Chamaeleon I star-forming region (Cha~I). The most recent census of the substellar population in $\sigma$\,Ori, published by \citet{2007A&A...470..903C}, lists 34 objects with $0.015<M<0.08\,M_{\odot}$ and 12 with lower masses. In their mass scale, the planetary-mass objects have spectral types M9 or later, indicating that they exhibit lower temperatures (and thus lower masses) than the coolest objects in NGC~1333. The numbers of BDs and planemos in $\sigma$\,Ori are robust against the particular choice of evolutionary tracks and uncertainties in age, distance, and reddening. According to \citet{2007A&A...470..903C}, the census is expected to contain $\sim 6$ contaminating foreground dwarfs, $\sim 4$ of them in the planemo regime. Thus, the best estimate for the planemo vs. BD ratio in $\sigma$\,Ori is 8/32 or $25\pm ^9_8$\%. This is a lower limit, as the survey does not show any cut-off of the mass function down to $0.006\,M_{\odot}$, their completeness limit. According to recent adaptive optics observations in the core of $\sigma$\,Ori \citep{2009A&A...493..931B} the survey by \citet{2007A&A...470..903C} might miss a number of objects with planetary masses close to the bright central star.

In the ONC, the highly variable extinction makes an unbiased census difficult. In the largest spectroscopic survey in the substellar regime to date, \citet{2009MNRAS.392..817W} have confirmed $\sim 38$ objects with masses at or below the Hydrogen burning limit. According to their mass estimates, 17 objects in this sample have $0.015<M<0.08\,M_{\odot}$ and 20 objects have even lower masses. The masses are based on a comparison of the observed HR diagram with evolutionary tracks (including COND03 and BCAH98, as used in this paper) and thus rely partly on spectroscopic temperatures, comparable to our estimates in NGC~1333. Deriving the mass limits from the dereddened H-band photometry, as given by \citet{2009MNRAS.392..817W}, yields 15 objects with $0.015<M<0.08\,M_{\odot}$ and 14 below (assuming a distance of 450\,pc). Their analysis of the mass function thus supports the presence of a planetary-mass population in the ONC.

There are two alternative estimates for the frequency of planemos in the ONC in the literature: According to \citet{2006MNRAS.373L..60L}, 7.5\% (1-14\%) of the total population in the cluster have a mass of 0.003-0.015$\,M_{\odot}$, where the wide range of possible values is mostly a reflection for age uncertainty. \citet{2005MNRAS.361..211L} find a drop by factor two in the mass function at the Deuterium burning limit, a deviation from a flat mass function in the substellar regime. Their upper limit for the fraction of planemos in the total cluster population is 10-13\%. 

For Cha~I, the most up-to-date unbiased census by \citet{2007ApJS..173..104L} yields a total cluster population of 226 objects, among them 28 brown dwarfs, and four sources with masses below the Deuterium burning limit. Based on these numbers, the planemo fraction is 2\% relative to the total population and the planemo vs. brown dwarf ratio is 17\%. In this region, however, the census is complete only down to $\sim 0.01\,M_{\odot}$. The mass function is flat in the substellar regime and does not show indications for having reached the minimum mass \citep{2007ApJS..173..104L}. Thus, the total number of planemos may be twice as high, depending on the cut-off of the IMF. In fact, \citet{2008ApJ...684..654L} published one more source in the planetary-mass regime based on a Spitzer mid-infrared detection.

\begin{figure*}[t]
\center
\includegraphics[angle=-90,width=16cm]{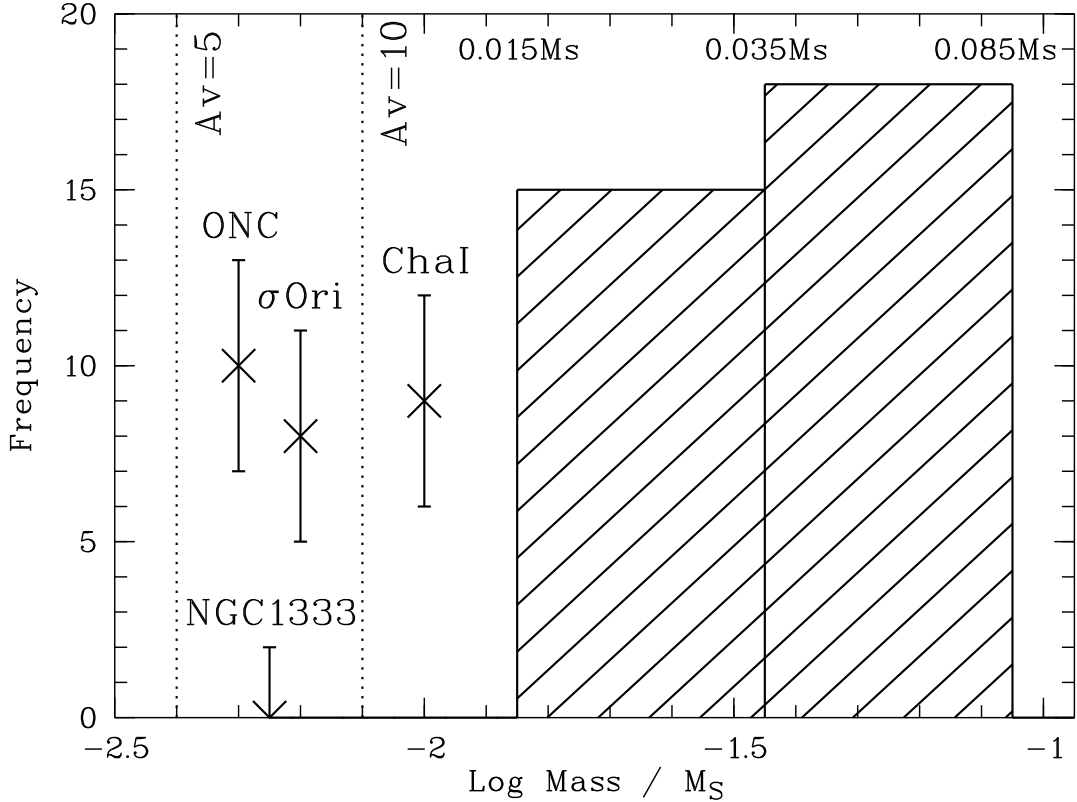} 
\caption{Mass distribution of brown dwarfs in NGC~1333 (hatched histogram) and the deficit of planetary-mass objects. The three labelled datapoints show the predicted number of planemos in NGC~1333 based on the surveys in $\sigma$\,Ori, the ONC, and in ChaI. Errorbars correspond to 1$\sigma$. The x-values for these three datapoints roughly correspond to the completeness limits in the literature surveys for these regions. The non-detection of planetary-mass objects points to a deficit of objects in this mass regime compared with other clusters. The dotted lines refer to our detection limit for $Av=5$ and 10\,mag..
\label{f10}}
\end{figure*}

Based on these literature findings, we can estimate the expected number of planemos in NGC~1333. This is illustrated in Fig.\ \ref{f10}. In our survey area, we find 19 brown dwarfs, seven of them previously identified by \citet{2004AJ....127.1131W}. In addition, the area covers 14 more substellar objects from other sources (see Sect.\ \ref{s51}). With one exception, all these brown dwarfs exhibit optical extinctions $<10$\,mag. For this extinction value, our survey is complete down to $\sim 0.008\,M_{\odot}$. The number of confirmed planetary-mass objects in the survey limits is zero. Based on the total sample covered in our spectroscopic observations, the 1$\sigma$ upper limit for this non-detection is about 2 objects. 

Assuming the planemo vs. brown dwarf ratio as seen in the $\sigma$\,Ori cluster, we would expect to find $8\pm 3$ planemos in NGC~1333, which has to be considered a lower limit, given the fact that the bottom of the mass function has not been found yet in $\sigma$\,Ori. The ONC survey by \citet{2009MNRAS.392..817W} predicts a number of planemos roughly comparable with the number of brown dwarfs, i.e.\ $\sim 33$ for NGC~1333. Based on the two other estimates, and assuming a total cluster population of $\sim 150$ in NGC~1333 \citep{2008ApJ...674..336G} we would expect $<20$, most likely $\sim 10$ planemos in the cluster. Based on the mass function in Cha~I, we expect 6-12 planemos in NGC~1333. Thus, all estimates for the expected number of planetary-mass objects are as high as or higher than the $2\sigma$ limit derived from our survey in NGC~1333. We conclude that NGC~1333 likely shows a deficit of planemos compared with all other star-forming regions where comparable survey depth has been reached.

We arrive at essentially the same conclusion when we use temperatures instead of masses: The coolest object in our survey has an effective temperature of 2500\,K, corresponding to a spectral type of M9 \citep{2003ApJ...593.1093L}. In $\sigma$\,Ori, there are 12 objects known with later spectral types \citep{2001A&A...377L...9B,2001ApJ...558L.117M}; a significant fraction of the low-mass population is not spectroscopically confirmed. The ONC survey by \citet{2009MNRAS.392..817W} gives 17 sources with $>$M9 and effective temperatures $<2500$\,K, about half of their total sample of substellar objects. In the ChaI census, the four objects with planetary masses all have spectral types later than M9. In summary, working with the maybe more robust effective temperatures instead of masses does not change the expected numbers of planetary-mass objects in any significant way.

Our result is similar to one of the conclusions by \citet{2007AJ....133.1321G}: Although they were sensitive to objects fainter than those surveyed previously by \citet{2004AJ....127.1131W}, they did not detect lower mass objects, only sources with higher extinction. We confirm this finding based on a much larger survey with a factor of $\sim 300$ more area coverage, encompassing the full extent of the cluster. This result is even more remarkable when we take into account that the number of brown dwarfs in NGC~1333 is unusually high, as pointed out in Sect.\ \ref{s52}. 

The best explanation for this outcome is a significant drop in the mass function in NGC~1333 around the Deuterium burning limit, corresponding to effective temperatures below 2500\,K in this cluster. Compared with other clusters like $\sigma$\,Ori and the ONC, the mass function of NGC~1333 does not extend down to planetary masses. As our survey is deeper than previous work in other regions and covers a substantial sample of objects with follow-up spectroscopy, this is the best evidence so far for a cut-off in the mass function at very low masses. It is plausible that our survey has detected the minimum mass limit for star formation in this particular cluster, at around 0.012-0.02$\,M_{\odot}$. 

With the currently available sample, this is a 2$\sigma$ result and thus needs further verification. If confirmed, it would point to strong environmental dependences in the mass function, particularly the mininum mass of star formation. In all other regions surveyed with comparable depth, the mass function extends well below 0.01$\,M_{\odot}$, while the spectrum is truncated at $\sim 0.012-0.02\,M_{\odot}$ in our target cluster NGC~1333. 

At this point it is not clear what could cause a change in the minimum mass by a factor of two or more. Compared with regions as diverse as $\sigma$\,Ori, the ONC, and ChaI, there is no evidence that NGC~1333 is different in any significant way in terms of initial conditions, dust properties, or metallicity. Based on the currently available census, the object density in NGC~1333 is slightly higher than in Cha~I and significantly higher than in $\sigma$\,Ori. While the number of low-mass members in all three regions is probably comparable and in the range of 150-250 objects, the radii of their distribution are 0.3\,pc in NGC\,1333, 0.5-0.7\,pc in Cha~I, and 3-5\,pc in $\sigma$\,Ori \citep{2004AJ....128.2316S,2007ApJS..173..104L,2008ApJ...674..336G}. \citet{2008MNRAS.389.1556B} recently suggested that brown dwarfs in clusters form through the gravitational fragmentation of infalling gas. This could explain the overabundance of substellar objects in the denser NGC\,1333, but would also predict a higher number of planetary-mass objects. On the other hand, the differences in density in those three regions are, as explained in Sect. \ref{s44}, most likely a consequence of progressing dynamical evolution and thus a difference in age, and not necessarily in initial conditions. Thus, a consistent interpretation of our findings in terms of the formation process is currently lacking.

\section{Conclusions}
\label{s6}

We present the first results of the SONYC project, short for `Substellar Objects in Nearby Young Clusters'. We have obtained deep optical and near-infrared observations of the NGC~1333 cluster using the Subaru 8-m telescope. From the photometry in the i', z', J and K band, we have selected brown dwarf candidates. Follow-up infrared spectroscopy, also obtained at Subaru, is used to verify their nature. In the following, we summarize the most important findings of this survey:
\begin{itemize}
\item{In our large-scale, optical$+$near-infrared imaging survey, we reach completeness limits of 24.7 in i'-band, 20.8 in J-band, and 18.0 in K-band. In terms of object masses for members of NGC~1333, this corresponds roughly to mass limits of 0.008$\,M_{\odot}$ for $A_V\la 10$\,mag and 0.004$\,M_{\odot}$ for $A_V \la 5$\,mag, based on the COND03 and DUSTY00 evolutionary tracks. The spatial coverage of our survey is 0.25\,sqdeg and includes all previously identified cluster members as well as the central cloud cores NGC~1333-A and -B.}
\item{From the optical photometry, we select 196 objects with i'--z' colors in the range as expected for substellar and planetary-mass cluster members, including a number of previously identified brown dwarfs. 36 of these objects are chosen for follow-up spectroscopy. This spectroscopic sample does not show any bias in spatial coverage or optical/near-infrared colors with respect to the full photometric candidate sample.}
\item{Using multi-object spectroscopy in H- and K-band, we confirm 28 objects close to the cluster center as late-type sources, 21 of them from the sample selected from photometry. This is based on water absorption features in the H-band. The extinctions, derived from the $J-K$ color, for the 28 objects cover the range from 0 to 13\,mag. Fitting model spectra to the observed ones, we find effective temperatures of 2500-3900\,K with a typical uncertainty of $\pm 200$\,K.}
\item{From the 28 confirmed late-type objects, 19 have effective temperatures of 3000\,K or lower (and spectral types M6 or later). Combined with the clear indications for youth, this classifies them as probable substellar members of NGC~1333. Of this subsample, seven have been previously identified by \citet{2004AJ....127.1131W}. The remaining nine objects in the confirmed sample are possible steller members, which likely exhibit significant contamination by field dwarfs/giants.}
\item{The confirmed objects are strongly clustered around the core of NGC~1333. The majority of them shows evidence for disks in Spitzer mid-infrared colors. Both findings additionally provide evidence for membership.}
\item{Combining our survey results with previous studies, the current census of spectroscopically confirmed brown dwarfs in NGC~1333 is 33. For comparison, the number of stellar members with masses between 0.08 and 1$\,M_{\odot}$ is in the range of 50. This indicates a clear overabundance of brown dwarfs in NGC~1333; the ratio of substellar to stellar members with masses below 1$\,M_{\odot}$ is $1.5 \pm 0.3$, lower by a factor of 2-5 than in all other previously surveyed regions.}
\item{The low-mass limit of the confirmed brown dwarfs in NGC~1333 is 0.012-0.02$\,M_{\odot}$, but the completeness limits are at significantly lower masses. Thus, although being able to do so, we do not find a population of planetary-mass objects in this cluster. Scaling from literature results in other regions ($\sigma$\,Ori, ONC, Cha~I), we would expect to find 8-10 `planemos', but we find none. This indicates a cut-off in the mass spectrum around the Deuterium burning limit in NGC~1333; we have possibly reached the minimum mass of the IMF.}
\item{If confirmed, these last two findings point to regional/environmental variations as a major factor in determining the number of very low-mass objects and the minimum mass of the IMF. It is particularly difficult to explain simultaneously the overabundance of brown dwarfs and the deficit of planetary-mass objects in NGC~1333. Thus, our results present a challenge for the current theoretical framework of star formation.}
\item{The mass spectrum in the brown dwarf and planetary mass regime necessarily relies on evolutionary models. Although we have made every effort to derive robust conclusions, we cannot get around the fact that evolutionary models lack reliable calibrations at the lowest masses. This is particularly true at young ages, where initial conditions may affect mass estimates, for example. Thus, the lack of objects with $T<2500$\,K in NGC~1333 could in principle be a sign of an unusual mass function or unusual initial conditions, or both. On the other hand, testing and calibrating the models require that we first identify objects at these masses and ages. Making relative comparisons between sub-stellar populations of different clusters could also help. Our SONYC project is an important step towards achieving these goals.}
\end{itemize}

\acknowledgments
The authors would like to thank the Subaru staff, specifically the supporting astronomer Ichi Tanaka, for assistance during the observations. We are grateful to Elaine Winston for making a preprint of her work available to us prior to publication. Our referee, Paul Clark, has provided a constructive and helpful report. The research was supported in part by grants from the Natural Sciences and Engineering Research Council (NSERC) of Canada and an Early Researcher Award from the province of Ontario to RJ.

\bibliography{aleksbib}

\begin{thebibliography}{87}
\expandafter\ifx\csname natexlab\endcsname\relax\def\natexlab#1{#1}\fi

\bibitem[{{Ahmic} {et~al.}(2007){Ahmic}, {Jayawardhana}, {Brandeker}, {Scholz},
  {van Kerkwijk}, {Delgado-Donate}, \& {Froebrich}}]{2007ApJ...671.2074A}
{Ahmic}, M., {Jayawardhana}, R., {Brandeker}, A., {Scholz}, A., {van Kerkwijk},
  M.~H., {Delgado-Donate}, E., \& {Froebrich}, D. 2007, \apj, 671, 2074

\bibitem[{{Allard} {et~al.}(2001){Allard}, {Hauschildt}, {Alexander},
  {Tamanai}, \& {Schweitzer}}]{2001ApJ...556..357A}
{Allard}, F., {Hauschildt}, P.~H., {Alexander}, D.~R., {Tamanai}, A., \&
  {Schweitzer}, A. 2001, \apj, 556, 357

\bibitem[{{Allen} {et~al.}(2004){Allen}, {Calvet}, {D'Alessio}, {Merin},
  {Hartmann}, {Megeath}, {Gutermuth}, {Muzerolle}, {Pipher}, {Myers}, \&
  {Fazio}}]{2004ApJS..154..363A}
{Allen}, L.~E., {Calvet}, N., {D'Alessio}, P., {Merin}, B., {Hartmann}, L.,
  {Megeath}, S.~T., {Gutermuth}, R.~A., {Muzerolle}, J., {Pipher}, J.~L.,
  {Myers}, P.~C., \& {Fazio}, G.~G. 2004, \apjs, 154, 363

\bibitem[{{Allers} {et~al.}(2007){Allers}, {Jaffe}, {Luhman}, {Liu}, {Wilson},
  {Skrutskie}, {Nelson}, {Peterson}, {Smith}, \&
  {Cushing}}]{2007ApJ...657..511A}
{Allers}, K.~N., {Jaffe}, D.~T., {Luhman}, K.~L., {Liu}, M.~C., {Wilson},
  J.~C., {Skrutskie}, M.~F., {Nelson}, M., {Peterson}, D.~E., {Smith}, J.~D.,
  \& {Cushing}, M.~C. 2007, \apj, 657, 511

\bibitem[{{Allers} {et~al.}(2006){Allers}, {Kessler-Silacci}, {Cieza}, \&
  {Jaffe}}]{2006ApJ...644..364A}
{Allers}, K.~N., {Kessler-Silacci}, J.~E., {Cieza}, L.~A., \& {Jaffe}, D.~T.
  2006, \apj, 644, 364

\bibitem[{{Andersen} {et~al.}(2008){Andersen}, {Meyer}, {Greissl}, \&
  {Aversa}}]{2008ApJ...683L.183A}
{Andersen}, M., {Meyer}, M.~R., {Greissl}, J., \& {Aversa}, A. 2008, \apjl,
  683, L183

\bibitem[{{Aspin} {et~al.}(1994){Aspin}, {Sandell}, \&
  {Russell}}]{1994A&AS..106..165A}
{Aspin}, C., {Sandell}, G., \& {Russell}, A.~P.~G. 1994, \aaps, 106, 165

\bibitem[{{Baraffe} {et~al.}(1998){Baraffe}, {Chabrier}, {Allard}, \&
  {Hauschildt}}]{1998A&A...337..403B}
{Baraffe}, I., {Chabrier}, G., {Allard}, F., \& {Hauschildt}, P.~H. 1998, \aap,
  337, 403

\bibitem[{{Baraffe} {et~al.}(2002){Baraffe}, {Chabrier}, {Allard}, \&
  {Hauschildt}}]{2002A&A...382..563B}
---. 2002, \aap, 382, 563

\bibitem[{{Baraffe} {et~al.}(2003){Baraffe}, {Chabrier}, {Barman}, {Allard}, \&
  {Hauschildt}}]{2003A&A...402..701B}
{Baraffe}, I., {Chabrier}, G., {Barman}, T.~S., {Allard}, F., \& {Hauschildt},
  P.~H. 2003, \aap, 402, 701

\bibitem[{{Barrado y Navascu{\'e}s} {et~al.}(2001){Barrado y Navascu{\'e}s},
  {Zapatero Osorio}, {B{\'e}jar}, {Rebolo}, {Mart{\'{\i}}n}, {Mundt}, \&
  {Bailer-Jones}}]{2001A&A...377L...9B}
{Barrado y Navascu{\'e}s}, D., {Zapatero Osorio}, M.~R., {B{\'e}jar}, V.~J.~S.,
  {Rebolo}, R., {Mart{\'{\i}}n}, E.~L., {Mundt}, R., \& {Bailer-Jones},
  C.~A.~L. 2001, \aap, 377, L9

\bibitem[{{Basri}(2000)}]{2000ARA&A..38..485B}
{Basri}, G. 2000, \araa, 38, 485

\bibitem[{{Basri} \& {Brown}(2006)}]{2006AREPS..34..193B}
{Basri}, G., \& {Brown}, M.~E. 2006, Annual Review of Earth and Planetary
  Sciences, 34, 193

\bibitem[{{Bate}(2005)}]{2005MNRAS.363..363B}
{Bate}, M.~R. 2005, \mnras, 363, 363

\bibitem[{{Bate}(2009)}]{2009MNRAS.392..590B}
---. 2009, \mnras, 392, 590

\bibitem[{{Bate} {et~al.}(2003){Bate}, {Bonnell}, \&
  {Bromm}}]{2003MNRAS.339..577B}
{Bate}, M.~R., {Bonnell}, I.~A., \& {Bromm}, V. 2003, \mnras, 339, 577

\bibitem[{{Belikov} {et~al.}(2002){Belikov}, {Kharchenko}, {Piskunov},
  {Schilbach}, \& {Scholz}}]{2002A&A...387..117B}
{Belikov}, A.~N., {Kharchenko}, N.~V., {Piskunov}, A.~E., {Schilbach}, E., \&
  {Scholz}, R.-D. 2002, \aap, 387, 117

\bibitem[{{Bertin} \& {Arnouts}(1996)}]{1996A&AS..117..393B}
{Bertin}, E., \& {Arnouts}, S. 1996, \aaps, 117, 393

\bibitem[{{Bessell} \& {Brett}(1988)}]{1988PASP..100.1134B}
{Bessell}, M.~S., \& {Brett}, J.~M. 1988, \pasp, 100, 1134

\bibitem[{{Bonnell} \& {Bate}(2006)}]{2006MNRAS.370..488B}
{Bonnell}, I.~A., \& {Bate}, M.~R. 2006, \mnras, 370, 488

\bibitem[{{Bonnell} {et~al.}(2008){Bonnell}, {Clark}, \&
  {Bate}}]{2008MNRAS.389.1556B}
{Bonnell}, I.~A., {Clark}, P., \& {Bate}, M.~R. 2008, \mnras, 389, 1556

\bibitem[{{Bonnell} {et~al.}(2007){Bonnell}, {Larson}, \&
  {Zinnecker}}]{2007prpl.conf..149B}
{Bonnell}, I.~A., {Larson}, R.~B., \& {Zinnecker}, H. 2007, in Protostars and
  Planets V, ed. B.~{Reipurth}, D.~{Jewitt}, \& K.~{Keil}, 149--164

\bibitem[{{Boss}(2001)}]{2001ApJ...551L.167B}
{Boss}, A.~P. 2001, \apjl, 551, L167

\bibitem[{{Bouy} {et~al.}(2009){Bouy}, {Hu{\'e}lamo}, {Mart{\'{\i}}n},
  {Marchis}, {Barrado Y Navascu{\'e}s}, {Kolb}, {Marchetti}, {Petr-Gotzens},
  {Sterzik}, {Ivanov}, {K{\"o}hler}, \& {N{\"u}rnberger}}]{2009A&A...493..931B}
{Bouy}, H., {Hu{\'e}lamo}, N., {Mart{\'{\i}}n}, E.~L., {Marchis}, F., {Barrado
  Y Navascu{\'e}s}, D., {Kolb}, J., {Marchetti}, E., {Petr-Gotzens}, M.~G.,
  {Sterzik}, M., {Ivanov}, V.~D., {K{\"o}hler}, R., \& {N{\"u}rnberger}, D.
  2009, \aap, 493, 931

\bibitem[{{Burgasser} {et~al.}(2007){Burgasser}, {Reid}, {Siegler}, {Close},
  {Allen}, {Lowrance}, \& {Gizis}}]{2007prpl.conf..427B}
{Burgasser}, A.~J., {Reid}, I.~N., {Siegler}, N., {Close}, L., {Allen}, P.,
  {Lowrance}, P., \& {Gizis}, J. 2007, in Protostars and Planets V, ed.
  B.~{Reipurth}, D.~{Jewitt}, \& K.~{Keil}, 427--441

\bibitem[{{Caballero} {et~al.}(2007){Caballero}, {B{\'e}jar}, {Rebolo},
  {Eisl{\"o}ffel}, {Zapatero Osorio}, {Mundt}, {Barrado Y Navascu{\'e}s},
  {Bihain}, {Bailer-Jones}, {Forveille}, \&
  {Mart{\'{\i}}n}}]{2007A&A...470..903C}
{Caballero}, J.~A., {B{\'e}jar}, V.~J.~S., {Rebolo}, R., {Eisl{\"o}ffel}, J.,
  {Zapatero Osorio}, M.~R., {Mundt}, R., {Barrado Y Navascu{\'e}s}, D.,
  {Bihain}, G., {Bailer-Jones}, C.~A.~L., {Forveille}, T., \& {Mart{\'{\i}}n},
  E.~L. 2007, \aap, 470, 903

\bibitem[{{Caballero} {et~al.}(2008){Caballero}, {Burgasser}, \&
  {Klement}}]{2008A&A...488..181C}
{Caballero}, J.~A., {Burgasser}, A.~J., \& {Klement}, R. 2008, \aap, 488, 181

\bibitem[{{Chabrier}(2003)}]{2003PASP..115..763C}
{Chabrier}, G. 2003, \pasp, 115, 763

\bibitem[{{Chabrier} {et~al.}(2000){Chabrier}, {Baraffe}, {Allard}, \&
  {Hauschildt}}]{2000ApJ...542..464C}
{Chabrier}, G., {Baraffe}, I., {Allard}, F., \& {Hauschildt}, P. 2000, \apj,
  542, 464

\bibitem[{{Cushing} {et~al.}(2005){Cushing}, {Rayner}, \&
  {Vacca}}]{2005ApJ...623.1115C}
{Cushing}, M.~C., {Rayner}, J.~T., \& {Vacca}, W.~D. 2005, \apj, 623, 1115

\bibitem[{{Cutri} {et~al.}(2003){Cutri}, {Skrutskie}, {van Dyk}, {Beichman},
  {Carpenter}, {Chester}, {Cambresy}, {Evans}, {Fowler}, {Gizis}, {Howard},
  {Huchra}, {Jarrett}, {Kopan}, {Kirkpatrick}, {Light}, {Marsh}, {McCallon},
  {Schneider}, {Stiening}, {Sykes}, {Weinberg}, {Wheaton}, {Wheelock}, \&
  {Zacarias}}]{2003tmc..book.....C}
{Cutri}, R.~M., {Skrutskie}, M.~F., {van Dyk}, S., {Beichman}, C.~A.,
  {Carpenter}, J.~M., {Chester}, T., {Cambresy}, L., {Evans}, T., {Fowler}, J.,
  {Gizis}, J., {Howard}, E., {Huchra}, J., {Jarrett}, T., {Kopan}, E.~L.,
  {Kirkpatrick}, J.~D., {Light}, R.~M., {Marsh}, K.~A., {McCallon}, H.,
  {Schneider}, S., {Stiening}, R., {Sykes}, M., {Weinberg}, M., {Wheaton},
  W.~A., {Wheelock}, S., \& {Zacarias}, N. 2003, {2MASS All Sky Catalog of
  point sources.} (The IRSA 2MASS All-Sky Point Source Catalog, NASA/IPAC
  Infrared Science Archive.~http://irsa.ipac.caltech.edu/applications/Gator/)

\bibitem[{{de Zeeuw} {et~al.}(1999){de Zeeuw}, {Hoogerwerf}, {de Bruijne},
  {Brown}, \& {Blaauw}}]{1999AJ....117..354D}
{de Zeeuw}, P.~T., {Hoogerwerf}, R., {de Bruijne}, J.~H.~J., {Brown}, A.~G.~A.,
  \& {Blaauw}, A. 1999, \aj, 117, 354

\bibitem[{{Enoch} {et~al.}(2006){Enoch}, {Young}, {Glenn}, {Evans}, {Golwala},
  {Sargent}, {Harvey}, {Aguirre}, {Goldin}, {Haig}, {Huard}, {Lange},
  {Laurent}, {Maloney}, {Mauskopf}, {Rossinot}, \&
  {Sayers}}]{2006ApJ...638..293E}
{Enoch}, M.~L., {Young}, K.~E., {Glenn}, J., {Evans}, II, N.~J., {Golwala}, S.,
  {Sargent}, A.~I., {Harvey}, P., {Aguirre}, J., {Goldin}, A., {Haig}, D.,
  {Huard}, T.~L., {Lange}, A., {Laurent}, G., {Maloney}, P., {Mauskopf}, P.,
  {Rossinot}, P., \& {Sayers}, J. 2006, \apj, 638, 293

\bibitem[{{Folha} \& {Emerson}(1999)}]{1999A&A...352..517F}
{Folha}, D.~F.~M., \& {Emerson}, J.~P. 1999, \aap, 352, 517

\bibitem[{{Getman} {et~al.}(2002){Getman}, {Feigelson}, {Townsley}, {Bally},
  {Lada}, \& {Reipurth}}]{2002ApJ...575..354G}
{Getman}, K.~V., {Feigelson}, E.~D., {Townsley}, L., {Bally}, J., {Lada},
  C.~J., \& {Reipurth}, B. 2002, \apj, 575, 354

\bibitem[{{Greissl} {et~al.}(2007){Greissl}, {Meyer}, {Wilking}, {Fanetti},
  {Schneider}, {Greene}, \& {Young}}]{2007AJ....133.1321G}
{Greissl}, J., {Meyer}, M.~R., {Wilking}, B.~A., {Fanetti}, T., {Schneider},
  G., {Greene}, T.~P., \& {Young}, E. 2007, \aj, 133, 1321

\bibitem[{{Gutermuth} {et~al.}(2008){Gutermuth}, {Myers}, {Megeath}, {Allen},
  {Pipher}, {Muzerolle}, {Porras}, {Winston}, \& {Fazio}}]{2008ApJ...674..336G}
{Gutermuth}, R.~A., {Myers}, P.~C., {Megeath}, S.~T., {Allen}, L.~E., {Pipher},
  J.~L., {Muzerolle}, J., {Porras}, A., {Winston}, E., \& {Fazio}, G. 2008,
  \apj, 674, 336

\bibitem[{{Helling} {et~al.}(2008){Helling}, {Dehn}, {Woitke}, \&
  {Hauschildt}}]{2008ApJ...675L.105H}
{Helling}, C., {Dehn}, M., {Woitke}, P., \& {Hauschildt}, P.~H. 2008, \apjl,
  675, L105

\bibitem[{{Jordi} {et~al.}(2006){Jordi}, {Grebel}, \&
  {Ammon}}]{2006A&A...460..339J}
{Jordi}, K., {Grebel}, E.~K., \& {Ammon}, K. 2006, \aap, 460, 339

\bibitem[{{J{\o}rgensen} {et~al.}(2006){J{\o}rgensen}, {Harvey}, {Evans},
  {Huard}, {Allen}, {Porras}, {Blake}, {Bourke}, {Chapman}, {Cieza}, {Koerner},
  {Lai}, {Mundy}, {Myers}, {Padgett}, {Rebull}, {Sargent}, {Spiesman},
  {Stapelfeldt}, {van Dishoeck}, {Wahhaj}, \& {Young}}]{2006ApJ...645.1246J}
{J{\o}rgensen}, J.~K., {Harvey}, P.~M., {Evans}, II, N.~J., {Huard}, T.~L.,
  {Allen}, L.~E., {Porras}, A., {Blake}, G.~A., {Bourke}, T.~L., {Chapman}, N.,
  {Cieza}, L., {Koerner}, D.~W., {Lai}, S.-P., {Mundy}, L.~G., {Myers}, P.~C.,
  {Padgett}, D.~L., {Rebull}, L., {Sargent}, A.~I., {Spiesman}, W.,
  {Stapelfeldt}, K.~R., {van Dishoeck}, E.~F., {Wahhaj}, Z., \& {Young}, K.~E.
  2006, \apj, 645, 1246

\bibitem[{{Kirkpatrick} {et~al.}(2006){Kirkpatrick}, {Barman}, {Burgasser},
  {McGovern}, {McLean}, {Tinney}, \& {Lowrance}}]{2006ApJ...639.1120K}
{Kirkpatrick}, J.~D., {Barman}, T.~S., {Burgasser}, A.~J., {McGovern}, M.~R.,
  {McLean}, I.~S., {Tinney}, C.~G., \& {Lowrance}, P.~J. 2006, \apj, 639, 1120

\bibitem[{{Kroupa}(2001)}]{2001MNRAS.322..231K}
{Kroupa}, P. 2001, \mnras, 322, 231

\bibitem[{{Lada} {et~al.}(1996){Lada}, {Alves}, \&
  {Lada}}]{1996AJ....111.1964L}
{Lada}, C.~J., {Alves}, J., \& {Lada}, E.~A. 1996, \aj, 111, 1964

\bibitem[{{Lodieu} {et~al.}(2006){Lodieu}, {Hambly}, \&
  {Jameson}}]{2006MNRAS.373...95L}
{Lodieu}, N., {Hambly}, N.~C., \& {Jameson}, R.~F. 2006, \mnras, 373, 95

\bibitem[{{L{\'o}pez Mart{\'{\i}}} {et~al.}(2004){L{\'o}pez Mart{\'{\i}}},
  {Eisl{\"o}ffel}, {Scholz}, \& {Mundt}}]{2004A&A...416..555L}
{L{\'o}pez Mart{\'{\i}}}, B., {Eisl{\"o}ffel}, J., {Scholz}, A., \& {Mundt}, R.
  2004, \aap, 416, 555

\bibitem[{{Low} \& {Lynden-Bell}(1976)}]{1976MNRAS.176..367L}
{Low}, C., \& {Lynden-Bell}, D. 1976, \mnras, 176, 367

\bibitem[{{Lucas} \& {Roche}(2000)}]{2000MNRAS.314..858L}
{Lucas}, P.~W., \& {Roche}, P.~F. 2000, \mnras, 314, 858

\bibitem[{{Lucas} {et~al.}(2005){Lucas}, {Roche}, \&
  {Tamura}}]{2005MNRAS.361..211L}
{Lucas}, P.~W., {Roche}, P.~F., \& {Tamura}, M. 2005, \mnras, 361, 211

\bibitem[{{Lucas} {et~al.}(2006){Lucas}, {Weights}, {Roche}, \&
  {Riddick}}]{2006MNRAS.373L..60L}
{Lucas}, P.~W., {Weights}, D.~J., {Roche}, P.~F., \& {Riddick}, F.~C. 2006,
  \mnras, 373, L60

\bibitem[{{Luhman}(2007)}]{2007ApJS..173..104L}
{Luhman}, K.~L. 2007, \apjs, 173, 104

\bibitem[{{Luhman} {et~al.}(2005){Luhman}, {Adame}, {D'Alessio}, {Calvet},
  {Hartmann}, {Megeath}, \& {Fazio}}]{2005ApJ...635L..93L}
{Luhman}, K.~L., {Adame}, L., {D'Alessio}, P., {Calvet}, N., {Hartmann}, L.,
  {Megeath}, S.~T., \& {Fazio}, G.~G. 2005, \apjl, 635, L93

\bibitem[{{Luhman} \& {Muench}(2008)}]{2008ApJ...684..654L}
{Luhman}, K.~L., \& {Muench}, A.~A. 2008, \apj, 684, 654

\bibitem[{{Luhman} {et~al.}(2003){Luhman}, {Stauffer}, {Muench}, {Rieke},
  {Lada}, {Bouvier}, \& {Lada}}]{2003ApJ...593.1093L}
{Luhman}, K.~L., {Stauffer}, J.~R., {Muench}, A.~A., {Rieke}, G.~H., {Lada},
  E.~A., {Bouvier}, J., \& {Lada}, C.~J. 2003, \apj, 593, 1093

\bibitem[{{Mart{\'{\i}}n} {et~al.}(2001){Mart{\'{\i}}n}, {Zapatero Osorio},
  {Barrado y Navascu{\'e}s}, {B{\'e}jar}, \& {Rebolo}}]{2001ApJ...558L.117M}
{Mart{\'{\i}}n}, E.~L., {Zapatero Osorio}, M.~R., {Barrado y Navascu{\'e}s},
  D., {B{\'e}jar}, V.~J.~S., \& {Rebolo}, R. 2001, \apjl, 558, L117

\bibitem[{{Mathis}(1990)}]{1990ARA&A..28...37M}
{Mathis}, J.~S. 1990, \araa, 28, 37

\bibitem[{{McLean} {et~al.}(2003){McLean}, {McGovern}, {Burgasser},
  {Kirkpatrick}, {Prato}, \& {Kim}}]{2003ApJ...596..561M}
{McLean}, I.~S., {McGovern}, M.~R., {Burgasser}, A.~J., {Kirkpatrick}, J.~D.,
  {Prato}, L., \& {Kim}, S.~S. 2003, \apj, 596, 561

\bibitem[{{Miyazaki} {et~al.}(2002){Miyazaki}, {Komiyama}, {Sekiguchi},
  {Okamura}, {Doi}, {Furusawa}, {Hamabe}, {Imi}, {Kimura}, {Nakata}, {Okada},
  {Ouchi}, {Shimasaku}, {Yagi}, \& {Yasuda}}]{2002PASJ...54..833M}
{Miyazaki}, S., {Komiyama}, Y., {Sekiguchi}, M., {Okamura}, S., {Doi}, M.,
  {Furusawa}, H., {Hamabe}, M., {Imi}, K., {Kimura}, M., {Nakata}, F., {Okada},
  N., {Ouchi}, M., {Shimasaku}, K., {Yagi}, M., \& {Yasuda}, N. 2002, \pasj,
  54, 833

\bibitem[{{Mohanty} {et~al.}(2004){Mohanty}, {Jayawardhana}, \&
  {Basri}}]{2004ApJ...609..885M}
{Mohanty}, S., {Jayawardhana}, R., \& {Basri}, G. 2004, \apj, 609, 885

\bibitem[{{Moraux} {et~al.}(2007){Moraux}, {Bouvier}, {Stauffer}, {Barrado y
  Navascu{\'e}s}, \& {Cuillandre}}]{2007A&A...471..499M}
{Moraux}, E., {Bouvier}, J., {Stauffer}, J.~R., {Barrado y Navascu{\'e}s}, D.,
  \& {Cuillandre}, J.-C. 2007, \aap, 471, 499

\bibitem[{{Moraux} {et~al.}(2003){Moraux}, {Bouvier}, {Stauffer}, \&
  {Cuillandre}}]{2003A&A...400..891M}
{Moraux}, E., {Bouvier}, J., {Stauffer}, J.~R., \& {Cuillandre}, J.-C. 2003,
  \aap, 400, 891

\bibitem[{{Muench} {et~al.}(2007){Muench}, {Lada}, {Luhman}, {Muzerolle}, \&
  {Young}}]{2007AJ....134..411M}
{Muench}, A.~A., {Lada}, C.~J., {Luhman}, K.~L., {Muzerolle}, J., \& {Young},
  E. 2007, \aj, 134, 411

\bibitem[{{Nakajima} {et~al.}(1995){Nakajima}, {Oppenheimer}, {Kulkarni},
  {Golimowski}, {Matthews}, \& {Durrance}}]{1995Natur.378..463N}
{Nakajima}, T., {Oppenheimer}, B.~R., {Kulkarni}, S.~R., {Golimowski}, D.~A.,
  {Matthews}, K., \& {Durrance}, S.~T. 1995, \nat, 378, 463

\bibitem[{{Oasa} {et~al.}(2008){Oasa}, {Tamura}, {Sunada}, \&
  {Sugitani}}]{2008AJ....136.1372O}
{Oasa}, Y., {Tamura}, M., {Sunada}, K., \& {Sugitani}, K. 2008, \aj, 136, 1372

\bibitem[{{Ouchi} {et~al.}(2004){Ouchi}, {Shimasaku}, {Okamura}, {Furusawa},
  {Kashikawa}, {Ota}, {Doi}, {Hamabe}, {Kimura}, {Komiyama}, {Miyazaki},
  {Miyazaki}, {Nakata}, {Sekiguchi}, {Yagi}, \& {Yasuda}}]{2004ApJ...611..660O}
{Ouchi}, M., {Shimasaku}, K., {Okamura}, S., {Furusawa}, H., {Kashikawa}, N.,
  {Ota}, K., {Doi}, M., {Hamabe}, M., {Kimura}, M., {Komiyama}, Y., {Miyazaki},
  M., {Miyazaki}, S., {Nakata}, F., {Sekiguchi}, M., {Yagi}, M., \& {Yasuda},
  N. 2004, \apj, 611, 660

\bibitem[{{Padoan} \& {Nordlund}(2002)}]{2002ApJ...576..870P}
{Padoan}, P., \& {Nordlund}, {\AA}. 2002, \apj, 576, 870

\bibitem[{{Padoan} \& {Nordlund}(2004)}]{2004ApJ...617..559P}
---. 2004, \apj, 617, 559

\bibitem[{{Pickles}(1998)}]{1998PASP..110..863P}
{Pickles}, A.~J. 1998, \pasp, 110, 863

\bibitem[{{Preibisch}(2003)}]{2003A&A...401..543P}
{Preibisch}, T. 2003, \aap, 401, 543

\bibitem[{{Rebolo} {et~al.}(1995){Rebolo}, {Zapatero-Osorio}, \&
  {Martin}}]{1995Natur.377..129R}
{Rebolo}, R., {Zapatero-Osorio}, M.~R., \& {Martin}, E.~L. 1995, \nat, 377, 129

\bibitem[{{Rees}(1976)}]{1976MNRAS.176..483R}
{Rees}, M.~J. 1976, \mnras, 176, 483

\bibitem[{{Ridge} {et~al.}(2006){Ridge}, {Di Francesco}, {Kirk}, {Li},
  {Goodman}, {Alves}, {Arce}, {Borkin}, {Caselli}, {Foster}, {Heyer},
  {Johnstone}, {Kosslyn}, {Lombardi}, {Pineda}, {Schnee}, \&
  {Tafalla}}]{2006AJ....131.2921R}
{Ridge}, N.~A., {Di Francesco}, J., {Kirk}, H., {Li}, D., {Goodman}, A.~A.,
  {Alves}, J.~F., {Arce}, H.~G., {Borkin}, M.~A., {Caselli}, P., {Foster},
  J.~B., {Heyer}, M.~H., {Johnstone}, D., {Kosslyn}, D.~A., {Lombardi}, M.,
  {Pineda}, J.~E., {Schnee}, S.~L., \& {Tafalla}, M. 2006, \aj, 131, 2921

\bibitem[{{Salpeter}(1955)}]{1955ApJ...121..161S}
{Salpeter}, E.~E. 1955, \apj, 121, 161

\bibitem[{{Schmidt} {et~al.}(2008){Schmidt}, {Neuh{\"a}user}, {Seifahrt},
  {Vogt}, {Bedalov}, {Helling}, {Witte}, \& {Hauschildt}}]{2008A&A...491..311S}
{Schmidt}, T.~O.~B., {Neuh{\"a}user}, R., {Seifahrt}, A., {Vogt}, N.,
  {Bedalov}, A., {Helling}, C., {Witte}, S., \& {Hauschildt}, P.~H. 2008, \aap,
  491, 311

\bibitem[{{Sherry} {et~al.}(2004){Sherry}, {Walter}, \&
  {Wolk}}]{2004AJ....128.2316S}
{Sherry}, W.~H., {Walter}, F.~M., \& {Wolk}, S.~J. 2004, \aj, 128, 2316

\bibitem[{{Smith} {et~al.}(2002){Smith}, {Tucker}, {Kent}, {Richmond},
  {Fukugita}, {Ichikawa}, {Ichikawa}, {Jorgensen}, {Uomoto}, {Gunn}, {Hamabe},
  {Watanabe}, {Tolea}, {Henden}, {Annis}, {Pier}, {McKay}, {Brinkmann}, {Chen},
  {Holtzman}, {Shimasaku}, \& {York}}]{2002AJ....123.2121S}
{Smith}, J.~A., {Tucker}, D.~L., {Kent}, S., {Richmond}, M.~W., {Fukugita}, M.,
  {Ichikawa}, T., {Ichikawa}, S.-i., {Jorgensen}, A.~M., {Uomoto}, A., {Gunn},
  J.~E., {Hamabe}, M., {Watanabe}, M., {Tolea}, A., {Henden}, A., {Annis}, J.,
  {Pier}, J.~R., {McKay}, T.~A., {Brinkmann}, J., {Chen}, B., {Holtzman}, J.,
  {Shimasaku}, K., \& {York}, D.~G. 2002, \aj, 123, 2121

\bibitem[{{Stamatellos} \& {Whitworth}(2009)}]{2009MNRAS.392..413S}
{Stamatellos}, D., \& {Whitworth}, A.~P. 2009, \mnras, 392, 413

\bibitem[{{Stassun} {et~al.}(2006){Stassun}, {Mathieu}, \&
  {Valenti}}]{2006Natur.440..311S}
{Stassun}, K.~G., {Mathieu}, R.~D., \& {Valenti}, J.~A. 2006, \nat, 440, 311

\bibitem[{{Strom} {et~al.}(1976){Strom}, {Vrba}, \&
  {Strom}}]{1976AJ.....81..314S}
{Strom}, S.~E., {Vrba}, F.~J., \& {Strom}, K.~M. 1976, \aj, 81, 314

\bibitem[{{Testi} {et~al.}(2002){Testi}, {Natta}, {Oliva}, {D'Antona},
  {Comeron}, {Baffa}, {Comoretto}, \& {Gennari}}]{2002ApJ...571L.155T}
{Testi}, L., {Natta}, A., {Oliva}, E., {D'Antona}, F., {Comeron}, F., {Baffa},
  C., {Comoretto}, G., \& {Gennari}, S. 2002, \apjl, 571, L155

\bibitem[{{Walsh} {et~al.}(2007){Walsh}, {Myers}, {Di Francesco}, {Mohanty},
  {Bourke}, {Gutermuth}, \& {Wilner}}]{2007ApJ...655..958W}
{Walsh}, A.~J., {Myers}, P.~C., {Di Francesco}, J., {Mohanty}, S., {Bourke},
  T.~L., {Gutermuth}, R., \& {Wilner}, D. 2007, \apj, 655, 958

\bibitem[{{Weights} {et~al.}(2009){Weights}, {Lucas}, {Roche}, {Pinfield}, \&
  {Riddick}}]{2009MNRAS.392..817W}
{Weights}, D.~J., {Lucas}, P.~W., {Roche}, P.~F., {Pinfield}, D.~J., \&
  {Riddick}, F. 2009, \mnras, 392, 817

\bibitem[{{Whitworth} \& {Stamatellos}(2006)}]{2006A&A...458..817W}
{Whitworth}, A.~P., \& {Stamatellos}, D. 2006, \aap, 458, 817

\bibitem[{{Whitworth} \& {Zinnecker}(2004)}]{2004A&A...427..299W}
{Whitworth}, A.~P., \& {Zinnecker}, H. 2004, \aap, 427, 299

\bibitem[{{Wilking} {et~al.}(2004){Wilking}, {Meyer}, {Greene}, {Mikhail}, \&
  {Carlson}}]{2004AJ....127.1131W}
{Wilking}, B.~A., {Meyer}, M.~R., {Greene}, T.~P., {Mikhail}, A., \& {Carlson},
  G. 2004, \aj, 127, 1131

\bibitem[{{Winston} {et~al.}(2009){Winston}, {Megeath}, {Wolk}, {Hernandez},
  {Gutermuth}, {Muzerolle}, {Hora}, {Covey}, {Allen}, {Spitzbart}, {Peterson},
  {Myers}, \& {Fazio}}]{2009AJ....137.4777W}
{Winston}, E., {Megeath}, S.~T., {Wolk}, S.~J., {Hernandez}, J., {Gutermuth},
  R., {Muzerolle}, J., {Hora}, J.~L., {Covey}, K., {Allen}, L.~E., {Spitzbart},
  B., {Peterson}, D., {Myers}, P., \& {Fazio}, G.~G. 2009, \aj, 137, 4777

\bibitem[{{Yagi} {et~al.}(2002){Yagi}, {Kashikawa}, {Sekiguchi}, {Doi},
  {Yasuda}, {Shimasaku}, \& {Okamura}}]{2002AJ....123...66Y}
{Yagi}, M., {Kashikawa}, N., {Sekiguchi}, M., {Doi}, M., {Yasuda}, N.,
  {Shimasaku}, K., \& {Okamura}, S. 2002, \aj, 123, 66

\bibitem[{{Zapatero Osorio} {et~al.}(2000){Zapatero Osorio}, {B{\'e}jar},
  {Mart{\'{\i}}n}, {Rebolo}, {Barrado y Navascu{\'e}s}, {Bailer-Jones}, \&
  {Mundt}}]{2000Sci...290..103Z}
{Zapatero Osorio}, M.~R., {B{\'e}jar}, V.~J.~S., {Mart{\'{\i}}n}, E.~L.,
  {Rebolo}, R., {Barrado y Navascu{\'e}s}, D., {Bailer-Jones}, C.~A.~L., \&
  {Mundt}, R. 2000, Science, 290, 103

\end{thebibliography}

\clearpage

\end{document}